\documentclass[10pt]{article}
\usepackage{times}
\usepackage{amsmath,amsthm,amssymb,setspace,enumitem,epsfig, titlesec, verbatim,color,array,multirow,comment,graphicx,soul}
\usepackage{booktabs}
\usepackage[sort&compress,comma,round,numbers]{natbib}
\usepackage[bf,small]{caption}
\usepackage[export]{adjustbox}
\usepackage{rotating}
\usepackage[top=2.5cm,left=2.7cm,right=2.7cm,bottom=3.2cm]{geometry}
\usepackage{blkarray}
\usepackage[table]{xcolor}
\usepackage{xcolor}
\usepackage{mathtools}
\usepackage[
  colorlinks=false,
  pdfborder={0 0 0},
  hidelinks
]{hyperref}
\usepackage{setspace}  
\usepackage{tcolorbox}
\usepackage{caption}

\newcounter{extendedfigure}

\newenvironment{extendedfigure}[1][p]
{
    \refstepcounter{extendedfigure}
    \begin{figure}[#1]
    \captionsetup{name={Extended Data Figure}, labelformat=simple, labelsep=colon}
    
}
{
    \end{figure}
}

\definecolor{limegreen}{RGB}{50,205,50}      
\definecolor{darkgreen}{RGB}{0,100,0}        
\definecolor{brown}{RGB}{165,42,42}          
\definecolor{cornflowerblue}{RGB}{100,149,237} 
\definecolor{darkorange}{RGB}{255,140,0}     

\titleformat{\section}{\sffamily \fontsize{12}{12}\bfseries}{\thesection}{1em}{}
\titleformat{\subsection}{\sffamily \fontsize{10}{10.5}\bfseries}{\thesubsection}{1em}{}

\newtheoremstyle{plainCl1}
{9pt}
{15pt}
{\it}
{}
{\bfseries}
{.}
{2mm}
{}

\theoremstyle{plainCl1}

\usepackage{bbm}
\usepackage{color}

\allowdisplaybreaks[1]

\theoremstyle{remark}

\begin{document}
\title{\bfseries\sffamily \large When groups attract: coevolutionary dynamics of cooperation and individual- and group-based imitating rules}
\date{}
\author{\parbox[c]{16cm}{\centering \onehalfspacing \fontsize{11}{12}\selectfont 
Dini Wang$^{1,2,3,4,*}$,
Peng Yi$^{1,2}$,
Gang Yan$^{2,5}$,
Feng Fu$^{3,6,*}$\\[0.4cm]
$^1$College of Electronic and Information Engineering, Tongji University, Shanghai, China\\
$^2$Shanghai Research Institute for Intelligent Autonomous Systems, Tongji University, Shanghai, China\\
$^3$Department of Mathematics, Dartmouth College, Hanover, NH, USA\\
$^4$Program of Quantitative Social Science, Dartmouth College, Hanover, NH, USA\\
$^5$School of Physical Science and Engineering, Tongji University, Shanghai, China\\
$^6$Department of Biomedical Data Science, Geisel School of Medicine at Dartmouth,\\ Lebanon, NH, USA\\
$^*$\textbf{Correspondence author}: Dini Wang (diniwang1999@gmail.com); Feng Fu (fufeng@gmail.com)
}}

\maketitle
\onehalfspacing

\section*{Abstract}
\noindent 
Success-driven imitation is a fundamental aspect of social learning, yet group interactions beyond pairwise require more subtle imitating rules since  success can be observed at either the group or individual level. 
How such heuristic imitating rules based on these distinct levels of social information compete and coevolve with collective behavior, particularly cooperation, remains poorly understood. 
Here, we address this issue by studying the coevolutionary dynamics of cooperation in public goods games and individual- and group-based imitating rules on hypergraphs. 
We distinguish three rules according to whether payoff-biased imitation operates at the individual level, the group level, or both. 
Our results identify a mutual reinforcement between cooperation and group-biased imitation: preferentially learning from successful groups promotes cooperation, while cooperators in turn favor group-biased imitation. 
We confirm this synergy between cooperation and group-biased imitation across diverse synthetic and empirical higher-order network populations. 
We further show that the subtle scale and organization of group interactions critically shape this coevolution, with intermediate group sizes providing the greatest advantage for cooperation. 
Our results reveal how learning whom to imitate can jointly evolve to shape collective cooperation.

\section*{Introduction}
Humans acquire much of what they know and do by observing and interacting with others, a process called social learning~\cite{rendell2010copy,chibaokabe2024social}.
This heuristic process underlies the transmission of culture and, in turn, shapes the sophistication and prosperity of human societies.
Among the most fundamental ways of social learning is imitation.
How individuals imitate has been extensively identified in behavioral experiments, with strategies ranging from copying the majority to copying successful individuals or groups.
How alternative ways of imitation draw on social information at distinct scales to shape collective behavior is a fundamental question across the social, economic, and behavioral sciences.

Evolutionary game theory offers a powerful framework for studying the imitation dynamics of social traits, yielding a rich body of theoretical insights~\cite{nowak2006five}.
Of particular interest is the public goods game~\cite{hardin1968tragedy}, in which collective welfare are optimized when all players cooperate, yet each individual is better off defecting.
To alleviate this cooperation dilemma, network reciprocity has long been recognized as an effective mechanism. 
Different ways of using individual-level information for imitation can result in different evolutionary outcomes of cooperation through pairwise interactions on networks.
For example, for a considerable number of networks, cooperation is more advantageous when individuals preferentially imitate a more successful neighbor (called death-birth update) than when individuals randomly imitate a neighbor based on payoff comparison (called pair-comparison update).

In reality, humans interact not only in pairs but also in groups, from group coordination in crisis response~\cite{shirado2020collective,zhang2026evolution} to collective intelligence in problem solving~\cite{riedl2021quantifying,kameda2022information,saveski2021algorithmic}.
These group interactions can yield varying collective outcomes, such as team effectiveness~\cite{zhang2026evolution}, resource acquisition~\cite{gavrilets2014solution}, and task performance~\cite{riedl2021quantifying,king2011performance}, reflecting the success of the group as a whole.
Numerous studies have shown that group success can substantially influence the overall level of cooperation in group-based populations~\cite{traulsen2006multilevel,cooney2023evolutionary,kido2025empirical,yao2026payoff}.
Group success is correlated, yet not always aligned with individual success~\cite{civilini2021,yao2026payoff}.
Hence, imitation through group interactions warrants the separate treatment of individual-level and group-level information.

An extensive body of research, broadly framed in multi-level selection theory~\cite{traulsen2006multilevel,cooney2023evolutionary,yagoobi2023categorizing}, has investigated the effect of selection at individual and group levels, either separately or jointly, on evolutionary outcomes.
Recent theoretical work has shown that success-biased imitation at distinct levels can fundamentally alter the fate of cooperators, among which imitation biased by pure group success is most conducive to the evolution of cooperation~\cite{wang2025emergence}.
The advantage of copying successful groups has also received empirical support~\cite{kido2025empirical}.

Despite these advances, most studies assume that individuals are prescribed with an invariant learning heuristic throughout the life time.
In fact, individuals may not only differ in how to learn but also learn how to learn.
On the one hand, real experiments have revealed substantial heterogeneity in how human subjects respond to social information and adapt their behavior~\cite{molleman2014consistent,grujic2012consistent}; for example, some of them are success-based learners while others are frequency-based learners~\cite{vandenberg2015focus}.
Such variation is partly associated with individual attributes like personality and cognitive ability~\cite{burtenshaw2026individual,vostroknutov2018role}.
On the other hand, learning rules themselves can adjust in response to personal experience and peer influence.
For example, the prior experiment has shown that individuals preferentially follow the learning rule that has more reliably predicted rewards in their past experience~\cite{schultner2025feature}.
Despite some experimental efforts to examine the conflicts and emergence of distinct learning rules~\cite{rendell2010copy}, the theoretical underpinnings remain largely unexplored.

Instead of asking how one fixed imitation strategy affects population dynamics, we ask how imitating rules biased by multi-level success compete, coevolve, and shape collective cooperative behavior.
To describe group interactions, we resort to hypergraphs, where individuals belong to multiple groups simultaneously.
Depending on the level at which success bias operates, we consider three categories of imitating rules: group-and-individual-biased (GIB), group-biased (GB), and individual-biased (IB).
We use evolutionary game theory to establish a unified framework for studying coevolutionary dynamics between cooperation and individual- and group-based imitation through group interactions on hypergraphs, with exploration allowed.
Once the focal individual is determined to update, it will either learn both the social behavior and imitating rule from the role model selected by its current adopted rule, or stochastically explore a strategy combination.
Based on this setup, we provide a generic theoretical framework to calculate the mutation-selection equilibrium of all the potential strategy frequencies on any hypergraph.

Systematic analyses of a variety of hypergraph structures demonstrate that GB imitating rule can promote cooperation to the greatest extent, and cooperators preferentially favor GB imitation when three rules coevolve.
Furthermore, empirical observations from 16,469 random hypergraphs of size 100 extend this mutual reinforcement between cooperation and GB rule to a broader concordance: GB rule is most promoted by populations with a high proportion of cooperators in the equilibrium state.
We find that effective hypergraph anchoring can overcome the failure of well-mixed groups to promote cooperation, with intermediate scales of group interactions providing the greatest advantage to cooperation.
We finally present the heuristic strategy choice by Large Language Models (LLMs) as a complementary perspective.
Our work bridges network reciprocity and multilevel selection, two fundamental mechanisms for the evolution of cooperation, and highlights the pivotal role of group-biased imitation in fostering cooperative societies.

\section*{Model}

\noindent We model the coevolutionary dynamics of social behaviors and distinct imitating rules in a structured population within group interactions.
Each individual carries two types of traits: a social behavior for game interactions, and an imitating rule that specifies how the individual selects a role model to learn from during strategy update. 
Whenever an individual decides to adjust its strategy by imitating a role model, it will copy both the social behavior and the imitating rule of that role model.

In our model, each individual interacts in multiple (at least one) groups concurrently, and as a consequence, these groups can jointly shape its strategic learning of social behavior and imitating rule.
Such a multi-group structure can be mathematically represented as a hypergraph~\cite{alvarez2021}, where individuals reside on nodes interconnected by hyperedges.
The hyperedge, as a generalization of the edge in a classical graph, can accommodate a group of nodes, with its order, $g$, equal to the group size.
Analogous to the degree of a node in the graph, denoted as the number of its neighbors, we define the hyperdegree, $k$, as the number of hyperedges incident to a node in the hypergraph.
Compared to edges of order two that describe pairwise interactions, hyperedges of higher orders can faithfully capture group interactions, thereby enabling hypergraphs to be a powerful paradigm for studying higher-order collective dynamics among groups of interacting individuals~\cite{wang2026}.

In each successive time step, coevolutionary dynamics on a hypergraph unfold through game interactions followed by strategy update, as illustrated in Supplementary Fig.~S1.
Firstly, regarding game interactions, individuals play public goods games on hyperedges to which they belong.
In a public goods game on a hyperedge of order $g$, each player behaves either as a cooperator (C), who pays a cost into a public pool, or as a defector (D), who makes no contribution.
Without loss of generality, the cost is assumed to be $1$ hereafter.
The total investment from cooperators is multiplied by a synergy factor $r$ with $1 < r < g$, and then shared equally among all participants.
As a result, a cooperator receives the share of the gross benefit minus the cost, while a defector receives the benefit share alone.
When the group contains $n_{\mathrm{C}}$ cooperators, the payoffs to a cooperator and a defector are, respectively, given by
\begin{equation}
    f_{\mathrm{C}} = \frac{n_{\mathrm{C}}r}{g}-1,~~~
    f_{\mathrm{D}} = \frac{n_{\mathrm{C}}r}{g}.
\end{equation}

After games are played over all hyperedges simultaneously, each individual obtains its payoff by accumulating the payoffs from all games it participates in.
The group payoff is accordingly defined as the average individual payoff of all group members.
We then convert payoff into fitness using the exponential transformation $F = e^{\delta f}$, where $\delta > 0$ is the selection strength that determines how strongly the payoff differences affect the likelihood of being selected as a reference.
In what follows, we assume weak selection, that is $\delta  \ll 1$, in our theoretical analyses, while numerical simulations go beyond this regime.

Secondly, regarding strategy update, a focal individual is uniformly-at-random chosen to adjust its current strategy, including social behavior and imitating rule, by either imitation or mutation.
With the mutation rate $0 < v < 1$, the focal individual uniformly switches to a random strategy combination, including a social behavior and an imitating rule; 
otherwise, it identifies a role model in accordance with its own current imitating rule, and directly copies the strategy combination of the role model.

\begin{figure}[p]
    \centering
    \includegraphics[width=\linewidth]{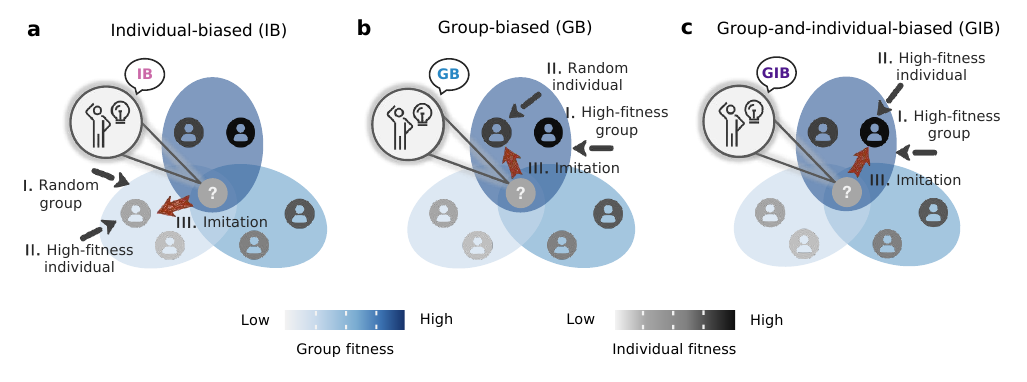}
    \vspace{0.02cm}
    \caption{\textbf{Group interactions on hypergraphs require nuanced social imitation processes that incorporate group-level and individual-level information.}
    The imitating process is defined on a hypergraph, where nodes represent individuals and hyperedges indicate groups.
    An focal individual, marked by `?', update its strategy in three steps: it first selects a group it belongs to, then selects an individual in that group as a role model, and finally imitates the role model for strategy updating (red arrows).
    Fitness bias can enter in the step of selecting a group, the step of selecting an individual, or both.
    Selecting a group and selecting an individual can be either biased by fitness, \textit{e.g.}, assigning a higher probability to a high-fitness group or individual, or neutral, \textit{e.g.}, assigning equal probability to every eligible group or individual.
    We therefore model three imitating rules: individual-biased (IB; \textbf{a}), group-biased (GB; \textbf{b}), and group-and-individual-biased (GIB; \textbf{c}).
    }  
    \label{fig:Fig1}
\end{figure}

Due to the multi-group feature of the hypergraph, we consider a three-stage imitation process within group interactions: the focal individual first selects a group to which it belongs, second selects an individual (including itself) in that group as the role model, and ultimately imitates that role model for a strategy update.
Both selecting a neighboring group and selecting an individual in that group can be either biased by fitness or purely neutral.
Quantitatively, biased selection means the probability of choosing an object, \textit{i.e.}, group or individual, is proportional to its fitness, while neutral selection gives all eligible objects an equal chance of being chosen.
Such a individual- and group-based imitation process produces three distinct imitating rules that differ in where fitness bias is adopted: individual-biased (IB; Fig.~\ref{fig:Fig1}a), group-biased (GB; Fig.~\ref{fig:Fig1}b), and group-and-individual-biased (GIB; Fig.~\ref{fig:Fig1}c).

Let us instantiate this imitation process on a hypergraph of size $N$, consisting of a node set $\mathcal{N}$ and a hyperedge set $\mathcal{E}$.
By multiplying the likelihoods of selecting a neighboring group and selecting an individual in that group, the probability that a randomly-chosen individual $j$ copies the strategy of its neighbor $i$ on this hypergraph under the imitating rules of IB, GB, and GIB can be respectively formulated as follows:
\begin{align}
e_{ij}^{\text{IB}}
&= \frac{1}{N} \sum_{I \in \mathcal{E}}
\underbrace{\frac{b(j,I)}{k_j}}_{\text{group-neutral}}
\cdot
\underbrace{\frac{b(i,I) F_{i}}
{\sum_{k \in \mathcal{N}} b(k,I) F_{k}}}_{\text{individual-biased}},
\label{rule-ib}
\\[8pt]
e_{ij}^{\text{GB}}
&= \frac{1}{N} \sum_{I \in \mathcal{E}}
\underbrace{\frac{b(j,I) F_{I}}
{\sum_{J \in \mathcal{E}} b(j,J) F_{J}}}_{\text{group-biased}}
\cdot
\underbrace{\frac{b(i,I)}{g_I}}_{\text{individual-neutral}},
\label{rule-gb}
\\[8pt]
e_{ij}^{\text{GIB}}
&= \frac{1}{N} \sum_{I \in \mathcal{E}}
\underbrace{\frac{b(j,I) F_{I}}
{\sum_{J \in \mathcal{E}} b(j,J) F_{J}}}_{\text{group-biased}}
\cdot
\underbrace{\frac{b(i,I) F_{i}}
{\sum_{k \in \mathcal{N}} b(k,I) F_{k}}}_{\text{individual-biased}},
\label{rule-gib}
\end{align}
where $b(i,I)=1$ indicates that node $i$ belongs to hyperedge $I$ and $0$ otherwise, and $F_i$ (and analogously $F_I$) is the fitness of individual $i$ (group $I$).
In doing so, the focal individual successfully accomplishes a strategy update.
This updating process is repeated over sufficient iterations until the system eventually converges to a stationary equilibrium.

\section*{General framework for mutation-selection equilibrium}
\noindent We explore how the three imitating rules differentially shape the long-term outcomes of evolution in structured populations, both when they evolve separately and when they coevolve.
By incorporating non-negligible mutation into imitation dynamics, the evolutionary process avoids fixation in which only one trait survives, and instead reaches a stationary equilibrium after sufficient interactions~\cite{alex2022_evaluating}. 
To this end, we establish a general theoretical framework for analyzing the mutation-selection equilibria on arbitrary hypergraphs, applicable to the coevolution of any subset of imitating rules considered here.
For a coevolving rule set $\mathcal{U}$ and an arbitrary mutation rate $v$, the stationary frequency 
of a target pair of behavior $s^* \in \{ \mathrm{C}, \mathrm{D} \}$ and rule $u^* \in \mathcal{U}$ is given under weak selection by
\begin{equation}
    \langle x_{(s^*,u^*)} \rangle = \frac{1}{2|\mathcal{U}|} + \frac{\delta (1-v)}{Nv} \sum_{u' \in \mathcal{U}} \left( \beta^{u'}_{(s^*,u^*)} r - \gamma^{u'}_{(s^*,u^*)} \right) + \mathcal{O}(\delta^2),
    \label{main_result}
\end{equation}
where $|\mathcal{U}|$ is the cardinality of the rule set.
Here, $\beta^{u'}_{(s^*,u^*)} r$ (resp. $\gamma^{u'}_{(s^*,u^*)}$) quantifies the effective benefit (resp. cost) induced by rule $u’$ on individuals with $(s^*,u^*)$; their explicit forms are offered in Methods and further detailed in Supplementary Information.
In other words, each coevolving rule contributes additively to the first-order deviation with respect to the selection strength from neutral equilibrium, \textit{i.e.}, $1/(2|\mathcal{U}|)$.
A special case is that when only a single rule operates, Eq.~(\ref{main_result}) would reduce to the stationary frequency of either cooperation or defection under that rule.

Of particular interest is the stationary frequency of cooperation in a finite population, denoted as $\langle x_{\mathrm{C}} \rangle$, which equals the accumulated proportions of strategy combinations associated with cooperation.
The mutation-selection balance is deemed to favor cooperation over defection if $\langle x_{\mathrm{C}} \rangle > 1/2$ (equivalently, the stationary frequency of defection $\langle x_{\mathrm{D}} \rangle < 1/2$), naturally yielding a critical synergy factor for cooperation, $r^*$, to demarcate the condition for when cooperation is favored over defection.
It is well-known that in a canonical public goods game, 
a larger group generally requires a greater synergy factor to relieve the cooperation dilemma.
For comparability amongst hypergraphs spanning a range of average hyperedge sizes (or orders), we instead define the rescaled critical synergy factor $R^*$ by normalizing $r^*$ with the average order for a specific hypergraph to measure its prospect of cooperation.
Typically, a smaller $R^*$ indicates a greater propensity for cooperation on a concrete hypergraph.
Beyond the condition for cooperation, no matter whether multiple rules evolve separately or coevolve, another crucial question unique to the coevolving case is how these rules compete and expand throughout cooperative populations.
We address it by means of the stationary frequency of the strategy combination associated with cooperation,  $\langle x_{(\mathrm{C}, u^*)} \rangle$ for every potential $u^*$.

\section*{Mutual reinforcement between cooperation and group-biased rule}
As an empirical illustration, we account for a coauthorship network capturing scientific collaborations through $24$ joint publications (hyperedges) among $71$ authors (nodes)~\cite{Benson2018}.
In this higher-order population, we study, both theoretically and experimentally, the propensities for cooperation under each single rule and under three coevolving rules.
When each rule evolves separately, theoretical predictions obtained from Eq.~\ref{main_result} have a great alignment with simulations across varying rescaled synergy factors $R$ (Fig.~\ref{fig:Fig2}a) and mutation rates $v$ (Fig.~\ref{fig:Fig2}b).
With the increment of rescaled synergy factors $R$, there exists a critical threshold for each rule (dashed lines in Fig.~\ref{fig:Fig2}a), over which cooperation is dominant over defection, that is $\langle x_\mathrm{C} \rangle > 1/2$.
Specifically, cooperation is promoted by the mutation-selection balance with the lowest critical rescaled synergy factor under GB rule, $R^*_{\mathrm{GB}}$, followed by GIB, $R^*_{\mathrm{GIB}}$, and then IB, $R^*_{\mathrm{IB}}$.
In other words, GB is the most conducive to the evolution of cooperation, while IB is the least conducive.
Similarly, as mutation rates change (Fig.~\ref{fig:Fig2}c), GB consistently yields the highest average level of cooperation, with GIB intermediate and IB lowest.
In a broader sense, prioritizing group fitness, or group success, can remarkably facilitate the proliferation of cooperators throughout higher-order populations.

When the three rules coevolve on the coauthorship network, our theory also predicts simulation outcomes accurately (scatters and dashed curves in Fig.~\ref{fig:Fig2}a--d).
In this case, the stationary frequency of cooperation lies within the spectrum spanned by its counterparts under single rules, and so does its critical rescaled synergy factor $R^*$ required for cooperation (Fig.~\ref{fig:Fig2}a--b).
It is reasonable because our theory of Eq.~\ref{main_result} has pointed out that the stationary strategy frequency is in essence a compromise amongst the effects of different coevolving rules.
More importantly, cooperators spontaneously favor GB rule for strategy updating, since $\langle x_{(\mathrm{C}, \mathrm{GB})} \rangle$ accounts for a substantial fraction of cooperators at equilibrium (Fig.~\ref{fig:Fig2}c--d).
Other two rules are nearly tied among cooperators, with GIB slightly exceeding IB (Fig.~\ref{fig:Fig2}c--d).
To complement this finding, we also explore the mutation-selection equilibrium when each pair of rules, \textit{i.e.}, IB and GB, GB and GIB, as well as GIB and IB, coevolves in the same population structure .
Among such pairwise competitions, cooperators are more inclined towards GB whenever it is present; in the absence of GB, however, cooperators relatively prefer GIB to IB (Extended Data Fig.~\ref{fig:FigExt1}).
Therefore, these coevolutionary outcomes collectively reveal that cooperators have an intrinsic preference for group-biased imitation.

Indeed, there emerges a fascinating mutual reinforcement: GB rule promotes the spread of cooperation, and cooperators in turn favor the prevalence of GB rule.
To assess its robustness across diverse interaction structures, we analyze the mutation-selection equilibria on $2,979$ synthetic hypergraphs of size $30$, covering variations in structural properties, heterogeneity of hyperdegree and order distributions, and multi-community organization.
Details of their generation are provided in Methods.
When each rule evolves alone, these synthetic hypergraphs demonstrate a significant ranking of critical rescaled synergy factors required for cooperation throughout all sampled synthetic hypergraphs: $R_{\mathrm{GB}}^* < R_{\mathrm{GIB}}^* < R_{\mathrm{IB}}^*$ (Fig.~\ref{fig:Fig2}e), thereby reflecting the advantageous effect of GB on the promotion of cooperation.
When three rules coevolve together, 
cooperators are most strongly associated with GB (Fig.~\ref{fig:Fig2}f), in marked contrast to defectors, among whom GB is least represented (Fig.~\ref{fig:Fig2}g).
This, in turn, mirrors the enhancing effect of cooperators on the prevalence of GB across these synthetic hypergraphs.
Consistent results are evidenced on $12,082$ small-scale hypergraphs of sizes $6$ to $8$ as well, which demonstrates that GB lowers the rescaled synergy factor required to favor cooperation below one in $99.69\%$ of structures, far exceeding GIB and IB, which do so in only $0.14\%$ and $0.02\%$ of structures, respectively (Extended Data Fig.~\ref{fig:FigExt2}).
Overall, our systematic analysis of a broad ensemble of hypergraphs further underscores the generality of the mutual reinforcement between cooperation and GB rule.

This compelling mutual reinforcement further motivates a natural question: whether it offers valuable insights into real human societies.
We then examine $24$ empirical higher-order populations, ranging in size from $31$ to $510$, across a variety of human interaction contexts: political collaboration, coauthorship, human contact, online social media, email communication, co-review, and Q\&A threads (see Methods for specifics of network descriptions and Extended Data Fig.~\ref{fig:FigExt3} for illustrations as well as summary statistics).
The mutual reinforcement between cooperation and GB persists in all of these empirical populations: GB is invariably the most conducive to cooperation owing to its lowest critical threshold required for cooperation (Fig.~\ref{fig:Fig2}h); and cooperators consistently give priority to the adoption of GB in a reasonable parameter choice, while defectors generally disfavor GB in the same parameter range (Fig.~\ref{fig:Fig2}i--j).
For example, across the five political populations, the mean critical threshold for cooperation is $8.26$-fold higher under GIB than under GB ($R^*_{\mathrm{GIB}}=0.846$ versus $R^*_{\mathrm{GB}}=0.103$), and $11.77$-fold higher under IB ($R^*_{\mathrm{IB}}=1.207$).
Another example occurs in the History coauthorship network, where cooperators are most strongly associated with GB, with $\langle x_{(\mathrm{C},\mathrm{GB})}\rangle=0.151$ compared with $\langle x_{(\mathrm{C},\mathrm{IB})}\rangle=0.128$ and $\langle x_{(\mathrm{C},\mathrm{GIB})}\rangle=0.127$.
More importantly, our general framework offers a fair basis for comparing strategic evolution across populations in similar contexts, because analogous behavior patterns make the same model setup valid to a comparable extent.

\section*{Intuitions of mutual reinforcement between cooperation and GB rule}
To shed light on the rationale behind the mutual reinforcement between cooperation and GB rule, we start with some intuitions (Extended Data Fig.~\ref{fig:FigExtss2}).
When each single rule evolves individually, we disentangle the effects of the three imitating rules along selecting a group and an individual: group-biased versus group-neutral, and individual-biased versus individual-neutral.
In public goods games, defectors typically obtain higher individual payoffs compared to cooperators, but groups enriched with more cooperators often achieve higher group average payoffs than groups with more defectors.
Thus, group-biased selection is apt to groups with a higher abundance of cooperators, owing to its advantageous group average payoff, while group-neutral selection is not.
In contrast, individual-biased selection tends to favor defectors, while individual-neutral selection lacks this tendency and thereby comparatively favors cooperators.
The GB rule integrates group-biased selection with individual-neutral selection, and this combination accounts for its capacity to facilitate the evolution of cooperation.

When three rules coevolve simultaneously, GB rule stabilizes cooperation by making current cooperators more likely to remain cooperative, while IB and GIB rules relatively destabilize cooperation by making cooperators more prone to switch to defection.
As a result, the frequencies $\langle x_{(\mathrm{C},\mathrm{IB})} \rangle$ and $\langle x_{(\mathrm{C},\mathrm{GIB})} \rangle$ decline relative to $\langle x_{(\mathrm{C},\mathrm{GB})} \rangle$ in the population.
Conversely, regarding defectors, IB and GIB favor the maintenance of defection, while GB increases the likelihood of transitioning to cooperation.
This leads to the decrement of $\langle x_{(\mathrm{D},\mathrm{GB})} \rangle$ compared to $\langle x_{(\mathrm{D},\mathrm{IB})} \rangle$ and $\langle x_{(\mathrm{D},\mathrm{GIB})} \rangle$.
In sharp contrast to defectors, cooperators provide a favorable niche for the reproduction of GB rule.

\section*{Mechanism of evolution of cooperation under distinct imitating rules}
The aforementioned intuitions offer a qualitative interpretation of the mutual reinforcement between cooperation and GB rule.
To quantitatively understand the mechanism underlying the evolution of cooperation, particularly under three coevolving rules, we derive a mathematically rigorous condition for when cooperation is favored on any hypergraph, \textit{e.g.}, $\langle x_{\mathrm{C}} \rangle > 1/2$, as formulated in Fig.~\ref{fig:Fig3}a.
This condition can be understood as a global aggregation over individuals with local contributions scaled by their corresponding mutation-weighted reproductive values, $\pi^{\mathrm{mut}}$.
The mathematical definition of the mutation-weighted reproductive value can be found in Methods.
The effective contribution of individual $j$ depends on its probabilistic adoption of imitating rules on neutral drift.
To clarify these rule-specific terms, we highlight the critical positions (individuals) in an illustrative hypergraph (Fig.~\ref{fig:Fig3}b), and report simulation outcomes on a lattice hypergraph, where each node belongs to two hyperedges and each hyperedge has four nodes, in Fig.~\ref{fig:Fig3}c--d.
Conditioned on $j$ adopting GB rule, its local contribution equals the difference between the average payoff of a neighboring hyperedge containing a given cooperator, $\bar{f}_I$ (Fig.~\ref{fig:Fig3}c-2), and  the average payoff of a random neighboring hyperedge, $\bar{f}_J$ (Fig.~\ref{fig:Fig3}c-3); see Fig.~\ref{fig:Fig3}d-2.
Because hyperedges containing cooperators generally attain higher average payoffs, the GB term is more likely to be positive.
When $j$ uses IB rule, its local contribution compares the average payoff of a neighboring cooperator $\bar{f}_i$ (Fig.~\ref{fig:Fig3}c-1) with the average payoff of the shared hyperedge $\bar{f}_I$ (Fig.~\ref{fig:Fig3}c-2) involving both $i$ and $j$; see Fig.~\ref{fig:Fig3}d-1.
As cooperators typically bear the costs, their individual payoffs often fall below the group average payoff, rendering the IB term more likely to be negative.
The GIB term is equivalent to GB term plus IB term, 
hence it has an intermediate probability of being positive; see Fig.~\ref{fig:Fig3}d-3.

Through the lens of average payoffs, there is no doubt that GB contributes more positively to the promotion of cooperation in coevolutionary dynamics, compared to the other two.
Such rule-specific contributions to the condition for cooperation are embedded in local interaction structures, yet how population structures reshape the interplay between cooperation and imitating rules is still ambiguous.
To reveal the impact of hypergraph properties on the condition for cooperation, we consider a homogeneous hypergraph of size $N$, where each node belongs to $k$ hyperedges, each hyperedge contains $g$ nodes, and any pair of nodes can share at most one hyperedges.
Subsequently, we resort to the mean-field approximation by neglecting the dynamic correlations between strategies and rules among neighboring individuals.
In the extreme circumstance where every focal individual consistently adopts a single rule (alternatively, under each single rule), we derive the critical rescaled synergy factor for cooperation respectively as
\begin{align}
    &R_{\mathrm{GB}}^*
    =
    \frac{k(N-g)}
    {N(k+g-1)-kg^2},
    \\[5pt]
    &R_{\mathrm{IB}}^*
    =
    \frac{N-1}{N-g},
    \\[5pt]
    &R_{\mathrm{GIB}}^*
    =
    \frac{kN(g+1)-kg}
    {kN(g+1)+N(k+g-1)-2kg^2},
    \label{gib}
\end{align}
above which cooperation is promoted in the population.
The approximations on a lattice hypergraph match simulations very well, as shown in Fig.~\ref{fig:Fig3}d.
Supplementary approximations of critical thresholds of cooperation on a variety of synthetic hypergraphs show qualitative agreement with their corresponding exact solutions (Extended Data Fig.~\ref{fig:FigExt4}).
In this case, the critical rescaled synergy factor for cooperation for three coevolving rules is equal to that for GIB rule, that is $R^* = R_{\mathrm{GIB}}^*$.

To take this step further, $R^*$ (as well as $R^*_{\mathrm{GIB}}$) can be written as a convex combination of the other two thresholds:
$$R^* = R_{\mathrm{GIB}}^* = \theta \cdot R_{\mathrm{IB}}^* + (1-\theta) \cdot R_{\mathrm{GB}}^*,$$
where $\theta = {kg}/{(kg+k+g-1)}$, and $0 < \theta < 1$ for $k > 1$ and $g \geq 2$.
This formula has a natural explanation: GIB rule is fundamentally a sequential composition of group-biased and individual-biased selection, and the coevolution of three rules is a mixture of the outcomes of the pure rules.
Moreover, within this parameter range, we see that $\theta > 1/2$, which means $R^*$ (as well as $R^*_{\mathrm{GIB}}$) is more strongly shaped by $R_{\mathrm{IB}}^*$ than by $R_{\mathrm{GB}}^*$.

In particular, on relatively sparse homogeneous hypergraphs whose hyperdegree and order are far below the population size, \textit{i.e.}, $N \gg k$ and $N \gg g$, we have 
\begin{align}
    &R_{\mathrm{GB}}^*
    \rightarrow
    \frac{k}{k+g-1},
    \\[5pt]
    &R_{\mathrm{IB}}^*
    \rightarrow
    1,
    \\[5pt]
    &R^* = R_{\mathrm{GIB}}^*
    \rightarrow
    \frac{kg+k}{kg+k+g-1},
\end{align}
Typically, we say that a specific hypergraph is able to favor cooperation if the critical rescaled synergy factor for cooperation is smaller than $1$.
Therein, global cooperation can be favored, \textit{i.e.}, $\langle x_{\mathrm{C}} \rangle > 1/2$, even for some $R<1$ where local social dilemmas generally persist.
If $N \gg k$ and $N \gg g$, cooperation can be promoted under three coevolving rules because $R^* < 1$; likewise, when a single rule evolves, cooperation can be readily promoted by GB, thanks to its smallest estimated critical threshold for cooperation.
Through these explicit mathematical expressions, we quantitatively explain a universal mechanism of three coevolving rules in enhancing cooperation on hypergraphs, especially emphasizing the positive contribution of GB rule.

\section*{Intermediate group scale gives cooperation the greatest advantage}
We further ask: what kind of hypergraph structures can best promote the emergence of cooperation when three rules coevolve?
Of particular importance is the impact of the hyperedge size, namely order, on the mutation-selection equilibrium of cooperation, because order encodes the scale, \textit{i.e.}, number of individuals, of the higher-order interaction.
Substituting Eq.~(\ref{gib}) of $R_{\mathrm{GIB}}^*$ into $R^* = R_{\mathrm{GIB}}^*$, we then differentiate $R^*$ with respect to the order $g$, and find that an interior minimum exists conditioned on the hyperdegree $k < 2$.
Thus, for sufficiently large hypergraphs with $N \gg k$ and $k < 2$, the interior minimizer of $R^*$, when it lies in the feasible range of $g$, asymptotically satisfies 
$$ 
g \sim \sqrt{\frac{N(2-k)}{2k}} -1. 
$$
Putting this scaling into $R^*$ gives the corresponding asymptotic minimum threshold, $R^*_{\min} \sim {k}/({k+1})$, above which cooperation is promoted.
In general, when three rules coevolve, large sparse hypergraphs with intermediate hyperedge sizes could make the spread of cooperation easier.

This general principle does not extend to the well-mixed hypergraph, in which all nodes belong to one hyperedge.
In such a well-mixed hypergraph, cooperation in standard public goods games is never promoted under three coevolving rules.
This is because its stationary frequency of cooperation always stays below $1/2$ regardless of hyperedge (or hypergraph) size~(Fig.~\ref{fig:Fig4}b) and its critical threshold required for cooperation does not exist any more, as also shown by the explicit formula in Methods.
In this case, IB and GIB rules collapse to the imitation process well-known in pairwise local update, where the focal individual imitates a neighbor, including itself, with the probability proportional to that neighbor’s fitness; and GB rule reduces to neutral drift, where the focal individual randomly imitates a neighbor, including itself.
Intuitively, imitation updating tends to drive the population toward defection because defectors obtain higher payoffs in the group public goods game.
Neutral drift leaves the stationary frequency of cooperation unbiased at one half.
Together, the well-mixed hypergraph typically impede the evolution of cooperation when three rules coevolve.

Yet, anchoring multiple well-mixed hypergraphs can remarkably alter the fate of cooperation under three coevolving rules.
When $m$ peripheral groups of size $n$ are anchored by a common node (Fig.~\ref{fig:Fig4}c), called anchoring-node hypergraph, the stationary frequency of cooperation can exceed that of defection at $R=1$ across a region of $(m,n)$ configurations.
For each fixed $m$, we find an optimal hyperedge size $n$ that maximizes the stationary frequency of cooperation in the given parameter setting; 
and as $m$ increases, this optimum converges to a moderate group size of $n=7$ among choices of $n$ from $2$ to $50$ (Fig.~\ref{fig:Fig4}d--e).

When each of $m$ peripheral groups of size $n$ offers one node and these nodes are anchored by a common group (Fig.~\ref{fig:Fig4}f), called anchoring-group hypergraph, cooperation can dominate defection at $R=1$ especially for $n < 10$.
For each fixed $m$, there also exists an optimal hyperedge size $n$ that yields the maximal stationary frequency of cooperation in the given parameter range;
and with increasing $m$, this optimum transitions from $n=9$ to $n=5$ with varying $n$ from $2$ to $20$ (Fig.~\ref{fig:Fig4}g--h).

Intuitively, a single well-mixed hypergraph contains no competing group through which group-biased imitation can identify a more successful cooperative group.
After anchoring, however, multiple peripheral groups become comparable through a shared node or a shared group.
Because groups enriched with cooperators tend to attain higher group average payoffs, group-biased imitation makes the shared node or the members of the shared group more likely to learn from these cooperator-rich groups and thereby to adopt cooperation.
Once these shared positions become cooperative, they can transmit cooperation to other peripheral groups while also reinforcing the groups in which cooperators are already clustered.
Therefore, effective hypergraph anchoring has the potential to rescue collective cooperation under three coevolving rules, among which moderate hyperedge sizes give cooperation the greatest advantage.

\section*{Concordance between cooperation and GB rule across hypergraphs}
To explore how higher-order structures mediate the interplay among cooperation and three coevolving rules on a large scale, we compute analytical results of mutation-selection equilibria on four families of $16,469$ synthetic hypergraphs of size $100$, including uniform random hypergraphs (URHs), uniform non-random hypergraphs (UNHs), scale-free hypergraphs (SFHs), and community hypergraphs (ComHs), as shown in Fig.~\ref{fig:Fig5}.
Generations and parameters of these synthetic hypergraphs are detailed in Methods.

We first report distributions of stationary frequencies of three distinct imitating rules over this ensemble of hypergraphs in Fig.~\ref{fig:Fig5}a, with each dot corresponding to one hypergraph.
For hypergraphs in the GB-dom region of Fig.~\ref{fig:Fig5}a, that is, GB has the the highest stationary frequency among the three rules, its dominance is typically only marginal.
However, for hypergraphs outside the GB-dom region, that is, GB has the the lowest stationary frequency among the three rules, this frequency can fall well below the neutral equilibrium (\textit{i.e.}, $1/3$), as also shown in Fig.~\ref{fig:Fig5}b.
The reason is that GIB and IB are often positively coupled across hypergraphs; that is to say, they tend to be jointly enriched or jointly suppressed on a specific hypergraph. 
When both rules are enriched, they occupy a substantial fraction of the stationary distribution, leaving GB well below the neutral equilibrium.

Simultaneously, higher stationary frequencies of GB are associated with higher stationary frequencies of cooperation across these hypergraphs, and vice versa, as shown in Fig.~\ref{fig:Fig5}a.
The same relationship can also be found in Fig.~\ref{fig:Fig5}i--k: GB is positively associated with cooperation across hypergraphs (Spearman correlation $\rho_s=0.754$), while IB and GIB are negatively associated with cooperation ($\rho_s=-0.769$ and $-0.630$, respectively).
In other words, hypergraphs conducive to cooperation tend to facilitate the proliferation of GB, and in turn, hypergraphs conducive to GB likewise tend to promote the spread of cooperation.
As a complementary observation, the same positive correlation between cooperation and GB is also shown in $11,117$ small hypergraphs of size $8$ in Extended Data Fig.~\ref{fig:FigExt7}.

Furthermore, we study the impact of varying structural properties on mutation-selection equilibria for these four families of synthetic hypergraphs.
As the average order $\langle g \rangle$, or average hyperedge size, increases across hypergraphs, the stationary frequency of cooperation first rises slightly and then declines (Fig.~\ref{fig:Fig5}d), corroborating the argument that the intermediate hyperedge size offers the greatest advantage for cooperation across a large scale of hypergraphs.
More interestingly, the stationary frequency of GB changes with order in concert with cooperation, while those of IB and GIB accordingly follow a complementary trend (Fig.~\ref{fig:Fig5}e), as frequencies of three rule always sum to one.
This reveals an intrinsic concordance between cooperation and GB rule in the variation of order across hypergraphs. 
Likewise, increasing the average hyperdegree $\langle k \rangle$ across hypergraphs also shows the concordance between cooperation and GB rule, whose stationary frequencies undergo a rise followed by a decline as well.

These empirical observations systematically reveal not only the positive correlation between cooperation and GB rule across hypergraphs, but also their concordant variation with hypergraph properties, \textit{i.e.}, order and hyperdegree.
Building on the mutual reinforcement between cooperation and the GB rule for a specific hypergraph, these results extend it to a broader perspective across a large scale of hypergraphs, among which the intermediate hyperedge size provides more opportunities for the evolution of both cooperation and GB rule.

\section*{Heuristic strategy choice by Large Language Models}
\noindent Finally, we explore how six leading Large Language Model (LLM) families, including Claude, Gemini, GPT, Llama, Gemma, and Mistral, make strategic decisions about imitating rules and game behaviors in a heuristic way.
Model adoption and parameter settings are offered in Methods, and specific prompts in this section are detailed in Supplementary Information.
We start with a baseline assessment of their preferences of three distinct imitating rules in the absence of additional conditions.
Using randomized pairwise comparisons, we quantify the pairwise preference by the win rate of every pair of rules, and evaluate the overall ranking of preference among three rules by Elo scores.
Elo scores are computed as detailed in Methods.
Averaged across six LLM models, the tournament among the three imitating rules shows that GIB achieves the highest score and a slightly higher win rate against the other two rules (Fig.~\ref{fig:Fig6}a).
Between IB and GB, IB has a modestly higher Elo score and win rate than GB (Fig.~\ref{fig:Fig6}a).

Next, we place the LLMs in the role of either a contributor (C) or a non-contributor (NC) in the public goods game, and test the effect of specific role contexts on their rule choices.
Here, we use contributor and non-contributor instead of cooperator and defector, to avoid introducing moral valence into the role labels.
By comparing Elo scores under role contexts to those in the baseline, we find that role contexts generally amplify the preference for GIB, especially when LLMs are assigned as contributors (Fig.~\ref{fig:Fig6}b). 
This suggests that when reasoning under social roles, LLMs continue to favor the information-rich rule that integrates both group- and individual-level considerations.

As a complementary counterpart, we then consider how LLMs adapt their game behaviors to changes in the rescaled synergy factor of public goods games, given a fixed social identity. 
Rather than presenting imitation rules in probabilistic forms, we recast them as intuitive social identities to simplify the decision context. 
Specifically, IB, GB, and GIB are represented as individual-oriented (IO), group-oriented (GO), and group-and-individual-oriented (GIO) identities, respectively.
As shown in Fig.~\ref{fig:Fig6}c, agents are, on average, least likely to choose contributors in public goods games when prompted with an IO identity, compared with GO and GIO.
In other words, the group-oriented identity leads to a larger propensity for acting as a contributor in group social dilemmas.

\section*{Discussion}
“The instinct of imitation is implanted in man from childhood, one difference between him and other animals being that he is the most imitative of living creatures,” Aristotle wrote in Poetics~\cite{aristotle1902poetics}.
Indeed, humans often imitate successful peers without even realizing it, which in turn renders imitation a common route for the spread of successful behavior, \textit{e.g.}, cooperative behavior, in human societies.
Yet imitation rarely takes place in pure peer-to-peer interactions.
People are typically embedded in multiple overlapped groups, so group success and individual success may both affect who becomes attractive to imitate.
For example, in scientific collaboration, a scientist could collaborate with different teams on several projects and publications. 
A productive research team may draw attention to its members, while an excellent scientist within that team may become a role model for collaborators in other projects. 

Evaluating whom to learn from within group interactions therefore demands both group-level and individual-level information.
One may prefer to learn from successful individuals, from successful groups, or from both sources of information.
These alternatives reflect individual-biased, group-biased, and group-and-individual-biased imitating rules, respectively.
Moreover, imitation need not stop at copying social behavior.
Individuals may copy how successful peers learn from their own role models as well, thereby adjusting the imitating rules they adopt over time.
Understanding how these rules coevolve with social behavior in higher-order populations can help clarify when group interactions facilitate the proliferation of cooperators in complex social systems.

We have modeled the coevolution of social behaviors and three distinct imitating rules within group interactions on hypergraphs.
Individuals might update both what they do and how they choose whom to imitate.
As such, a mutual reinforcement between cooperation and group-biased rule emerges: the group-biased imitating rule individually yields a lower critical threshold for cooperation to thrive, and cooperators in turn preferentially adopt the group-biased rule when three rules coevolve.
Across a variety of hypergraph structures, this relationship generalizes into a concordance between cooperation and group-biased imitation, that is, populations that promote cooperation to a larger extent tend to support a higher prevalence of the group-biased rule in the meanwhile.
We further find that effective hypergraph anchoring can markedly alter the fate of cooperators when three rules coevolve, among which intermediate scales of interactions typically give cooperators the greatest advantage for proliferation.
And so, group interactions on hypergraphs offer a broader perspective on the evolution of cooperation, highlighting group success as a catalyst for collective cooperation.

Our LLM experiments provide a complementary perspective on these imitating rules.
Although evolutionary dynamics on hypergraphs most strongly favor GB for the spread of cooperation, LLMs tend to prefer GIB when asked to choose among imitating rules.
This discrepancy is informative rather than contradictory.
Such a preference is plausible because LLMs are designed to synthesize diverse sources of information and optimize over candidate choices, rendering the group-and-individual-biased rule a natural heuristic for their strategic decisions.
Thus, AI reasoning offers an additional lens for interpreting the behavioral logic of distinct imitating patterns.

Our work offers several potential implications for the evolution of cooperation through the lens of group interactions.
The first is a charming mutual reinforcement between cooperation and group-biased imitation on hypergraphs.
Rather than merely focusing on how population structures shape reciprocal cooperation to evolve, our work points to updated rules by which cooperation can spread across the higher-order population.
In practice, social interventions, such as institutional designs~\cite{radzvilavicius2021adherence} or platform algorithms~\cite{matias2023influencing}, that amplify the visibility, reputation, or influence of successful groups may elevate the level of cooperation by encouraging individuals to learn from these groups.
Such interventions may engender a self-reinforcing loop: social interventions make successful groups salient as objects of imitation, group-biased imitation then helps cooperation spread, and cooperators become increasingly drawn to successful groups.
We therefore offer a mechanistic perspective on steering cooperation by publicizing successful groups without changing the underlying population structure.

The second implication is that strategic anchoring can help shift isolated defective groups to cooperation.
The failure of cooperation within isolated groups does not necessarily lead to the failure of cooperation in a connected population of groups.
For instance, cross-cultural migration or exchange programs can place individuals in more than one community, making them shared individuals that observe which groups achieve better collective outcomes~\cite{fu2012ingroup}.
Diaspora associations, workplaces or community organizations can similarly act as shared groups in which otherwise separated communities exchange experience.
And so, these shared individuals or shared groups break information barriers between isolated communities.
They allow successful groups to become visible as models, help their experience spread to other groups, and in turn reinforce the groups where successful practices have already accumulated.
In a broader sense, institutions may promote cooperation not only by changing incentives within isolated groups, but also by creating information-sharing bridges to transmit the strategy of successful groups.


Our work also connects to a theoretical gap in graph-structured meta-populations, where each node represents a well-mixed subpopulation, and these isolated groups are connected through migration.
Recent work categorized update mechanisms in which selection can act in the patch (group) level, in the individual level, or in both levels, and showed that many combinations remain unexplored~\cite{yagoobi2023categorizing}.
For example, in the death-birth process with migration coupled to death, the mechanisms $\mathrm{MddBb}$ and $\mathrm{MddbB}$ have no established analysis. 
These cases separate patch-level bias from individual-level bias when offspring are produced after a neutral death event.
Our work addresses an analogous problem in a more entangled setting.
On hypergraphs, an individual's fitness depends on the games from all groups in which it participates, rather than on a single local patch.
Group fitness, in turn, depends on the average fitness of its members, whose payoffs may also be influenced by other incident groups.
This interdependence makes group-level and individual-level information harder to separate than in graph-structured meta-populations.
In this sense, our analysis provides one route for addressing this gap, showing how group-level and individual-level biases can be separated, recombined and allowed to coevolve in overlapping higher-order populations.

Several limitations also point to future directions.
Firstly, our analytical framework relies on weak selection, so stronger selection may introduce nonlinear effects beyond first-order approximations.
Secondly, although the mutation-selection framework allows arbitrary mutation rates, the rare-mutation limit would place greater importance on the initial configuration of social behaviours and imitating rules across hypergraph positions, raising the question of which initial social arrangements are most conducive to cooperation.
Third, we treat the hypergraph structure as fixed when social behaviors and imitating rules coevolve. 
Allowing group memberships to form endogenously could reveal the feedback between network organization and social learning, and the stationary structure itself can also be part of evolutionary outcomes.
Fourth, we focus on standard public goods games, and other game models, such as coordination, trust, bargaining or punishment games, may alter when distinct imitating rules is most advantageous.
Fifth, behaviour and imitating rule are updated on the same evolutionary timescale in our model, but in real complex systems, learning rules reflect deeper cognitive biases, cultural heuristics, or institutional norms, and may nonetheless change more slowly than social behavior.
Finally, future work could study how human and artificial agents jointly adapt behaviours and update rules in human-AI hybrid systems.



\clearpage

\clearpage
    
\begin{figure}[p]
    \centering
    \includegraphics[width=\linewidth]{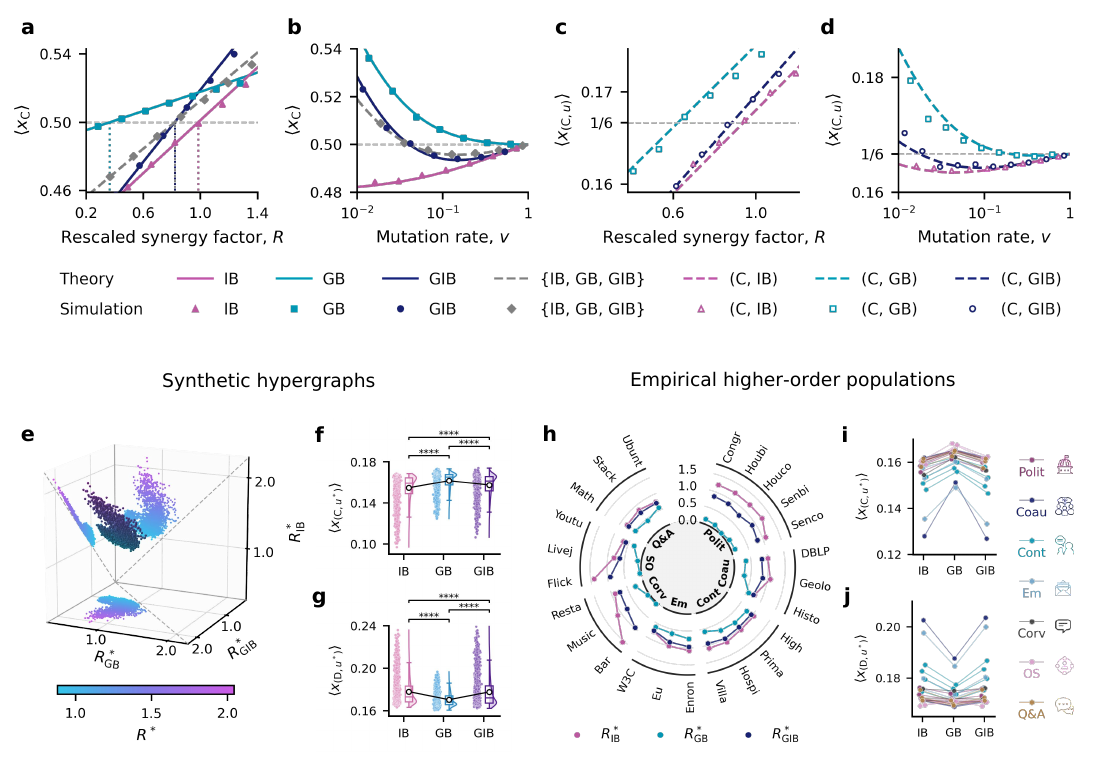}
    \vspace{0.02cm}
    \caption{\textbf{Cooperation and group-biased rule reinforce each other in higher-order populations.}
    \textbf{a}--\textbf{d}, Mutation-selection equilibria on a coauthorship higher-order network with $71$ authors (nodes) and $24$ publications (hyperedges).
    Scatters show simulations averaged over the last $10^8$ runs from $10^9$ time steps; 
    curves show theoretical predictions obtained from Eq.~\ref{main_result}.
    Gray horizontal dotted lines indicate neutral equilibria.
    \textbf{a}--\textbf{b}, Stationary frequencies of cooperation, $\langle x_{\mathrm{C}} \rangle$, as functions of the rescaled synergy factor $R$ (\textbf{a}) and the mutation rate $v$ (\textbf{b}), when each rule evolves separately (solid curves) and when three rules coevolve (dotted curves).
    Vertical dotted lines in \textbf{a} mark the critical rescaled synergy factors required for cooperation, at which $\langle x_{\mathrm{C}} \rangle=1/2$.
    \textbf{c}--\textbf{d}, Stationary frequencies of cooperation associated with the three rules when they coevolve.
    For \textbf{a}–\textbf{d}, model parameters are set to $\delta = 0.025$, $v = 0.05$ in \textbf{a} and \textbf{c}, and $R = 0.9$ in \textbf{b} and \textbf{d}.
    \textbf{e}--\textbf{g}, Analytical results across $2,979$ synthetic hypergraphs of size $30$.
    \textbf{e}, Critical rescaled synergy factors required for cooperation under each single rule ($R^*_{\mathrm{IB}}$, $R^*_{\mathrm{GB}}$, $R^*_{\mathrm{GIB}}$) and under three coevolving rules ($R^*$).
    \textbf{f}--\textbf{g}, Stationary frequencies of cooperation-rule pairs (\textbf{f}) and defection-rule pairs (\textbf{g}) under three coevolving rules.
    Box plots show the distributions (medians, interquartile ranges, and full ranges); 
    hollow markers indicate means, and points represent individual realizations.
    Asterisks denote statistical significance from Wilcoxon signed-rank tests, with $****~p < 10^{-4}$.
    \textbf{h}--\textbf{j}, Corresponding analytical results in $24$ empirical higher-order populations categorized into political (Polit), coauthorship (Coau), contact (Cont), email (Em), co-review (Corv), online social (OS), and question \& answer (Q\&A) networks.
    For \textbf{e}–\textbf{j}, model parameters are set to $v = 0.05$, $R = 1$, and $\delta = 0.01$.
    } 
    \label{fig:Fig2}
\end{figure}

\clearpage

\begin{figure}[p]
    \centering
    \includegraphics[width=0.9\linewidth]{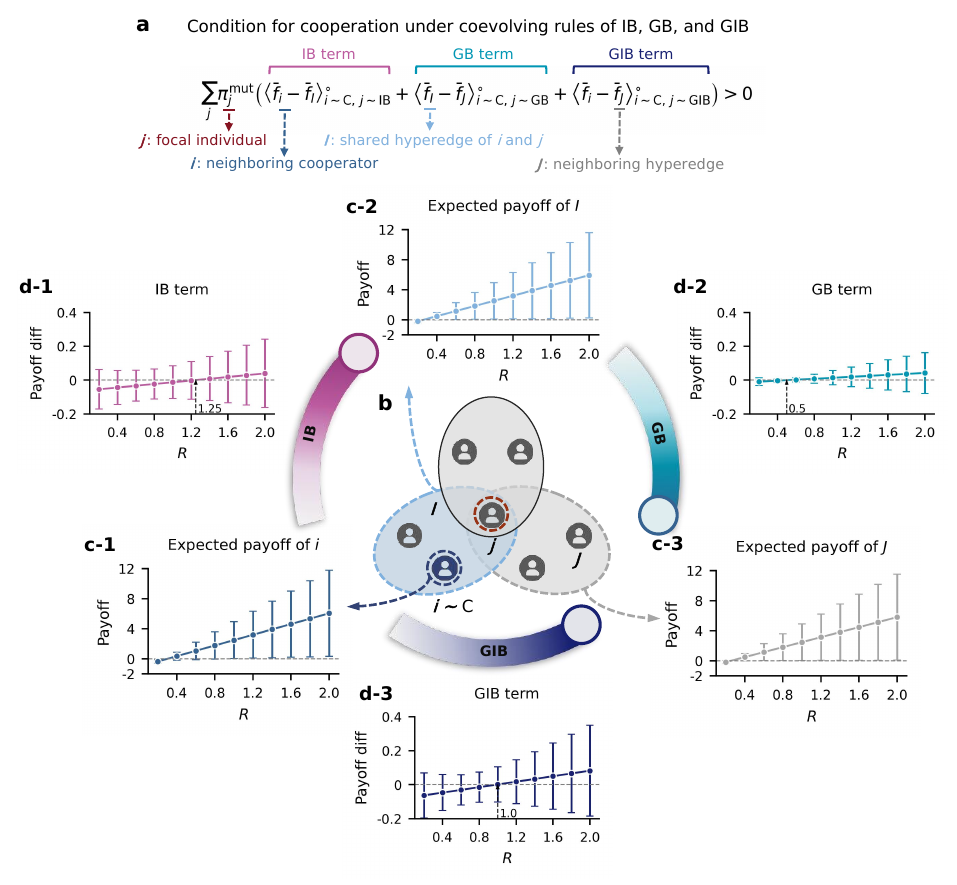}
    \vspace{0.3cm}
    \caption{
    \textbf{Understanding the coevolutionary dynamics of cooperation and imitating rules from an expected payoff perspective.}
    \textbf{a}, Mathematical condition for cooperation under coevolving IB, GB, and GIB rules, given by a population-level sum over individuals with contributions scaled by their mutation-weighted reproductive values, $\pi^{\text{mut}}$.
    For each focal individual $j$, its contribution can decompose into rule-specific terms determined by differences in expected payoffs among its neighboring cooperators $i$, $\bar{f}_i$, hyperedges $I$ shared by $i$ and $j$, $\bar{f}_I$, and neighboring hyperedges $J$ of $j$, $\bar{f}_J$, conditioned on different imitating rules as well as neutral drift ($\delta = 0$).
    \textbf{b}, Illustration of the local configuration around the focal individual $j$, depicting its interactions with $i$, $I$, and $J$ on the hypergraph.
    \textbf{c}--\textbf{d}, Expected payoffs of $i$ (\textbf{c-1}), $I$ (\textbf{c-2}), and $J$ (\textbf{c-3}), as well as rule-specific terms (\textbf{d-1}--\textbf{d-3}) as functions of the rescaled synergy factors $R$, centered on an arbitrary focal individual $j$ on a lattice hypergraph (size $16$, hyperdegree $2$, order $4$), with error bars indicating variability.
    Black dashed lines in \textbf{d} denote mean-field approximations of critical $R^*$ for the respective terms.
    Results are averaged over simulations of $10^9$ time steps with $v=10^{-4}$.
    } 
    \label{fig:Fig3}
\end{figure}

\clearpage

\begin{figure}[p]
    \centering
    \includegraphics[width=\linewidth]{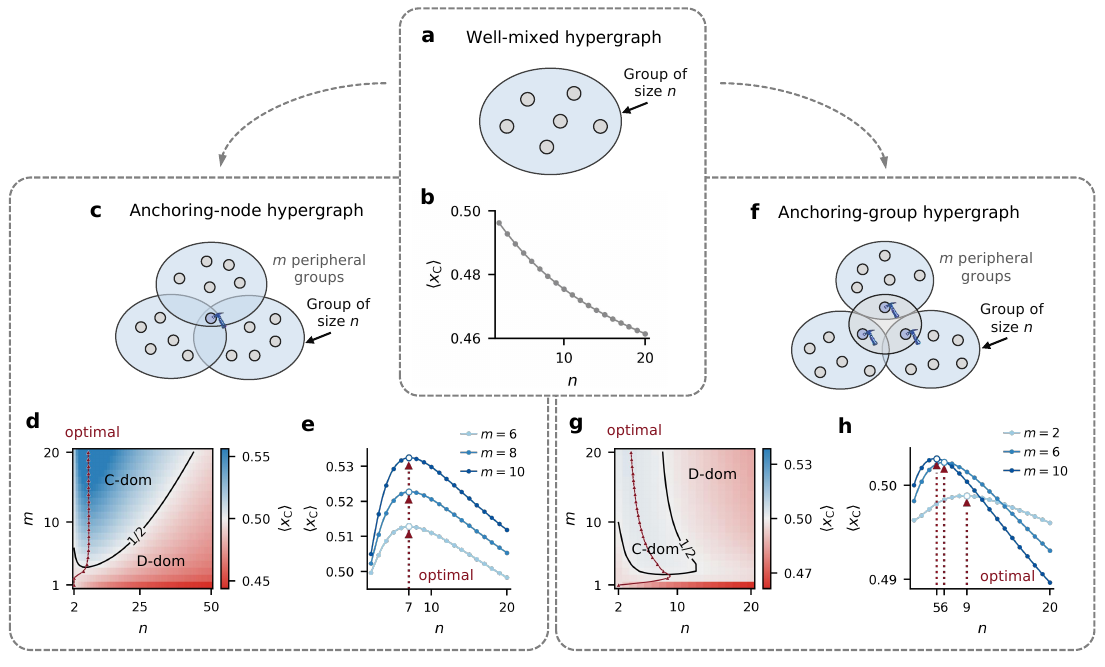}
    \vspace{0.3cm}
    \caption{\textbf{Anchoring well-mixed hypergraphs by a node or a group can rescue cooperation under three coevolving rules.}
    \textbf{a}, A well-mixed hypergraph composed of one group of size $n$.
    \textbf{b}, The stationary frequency of cooperation, $\langle x_{\mathrm{C}} \rangle$, decreases with respect to group size $n$ under three coevolving rules.
    \textbf{c}, A hypergraph connecting $m$ well-mixed populations of size $n$ by anchoring a node, called anchoring-node hypergraph.
    \textbf{d}, A heat map of theoretical stationary frequencies of cooperation, $\langle x_\mathrm{C} \rangle$, across group sizes ($n$) and numbers of peripheral groups ($m$) on anchoring-node hypergraphs. 
    The red line indicates the optimal $n$ for each row of $m$ (markers at integer $m$; interpolated between points),
    and the black curve delineates the boundary $\langle x_\mathrm{C} \rangle = 1/2$, separating cooperation-dominated (C-dom; $\langle x_\mathrm{C} \rangle > 1/2$) from defection-dominated (D-dom; $\langle x_\mathrm{C} \rangle < 1/2$) regimes.
    \textbf{e}, Analytical profiles of $\langle x_\mathrm{C} \rangle$ versus $n$, taken from cross-sections of \textbf{d} at fixed values of $m$.
    \textbf{f}, Hypergraphs connecting $m$ well-mixed populations of size $n$ by anchoring a group, called anchoring-group hypergraph.
    \textbf{g} and \textbf{h} are the counterparts of \textbf{d} and \textbf{e}, respectively, for anchoring-group hypergraphs.
    Parameters: $R=1$, $\delta = 0.025$, and $v = 0.05$.} 
    \label{fig:Fig4}
\end{figure}

\clearpage

\begin{figure}[htbp]
    \centering
    \includegraphics[width=\linewidth]{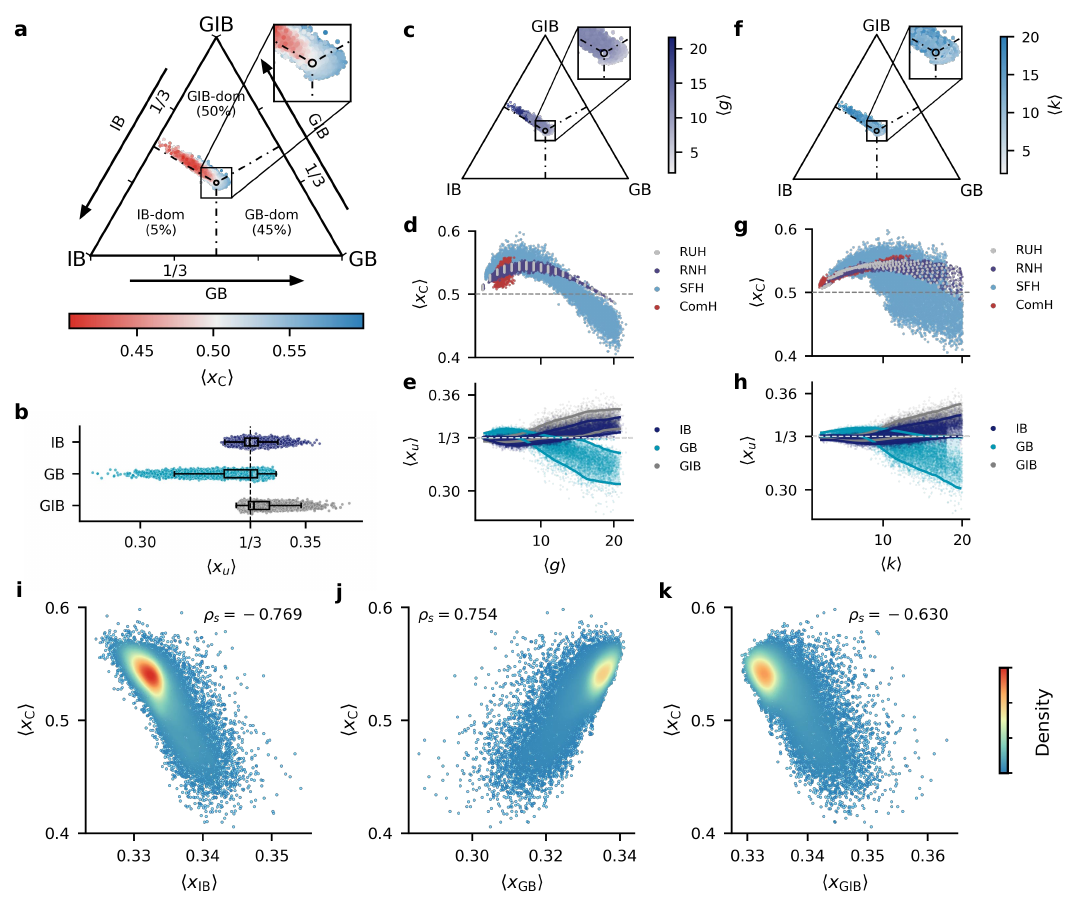}
    \vspace{0.3cm}
    \caption{\textbf{Analytical stationary strategy frequencies under three coevolving imitating rules across 16,469 synthetic hypergraphs of size 100.}
    All panels are based on the same ensemble of synthetic hypergraphs, including the random uniform hypergraph (RUH), the random non-uniform hypergraph (RNH), the scale-free hypergraph (SFH), and the community hypergraph (ComH).
    \textbf{a}, Ternary plot of IB, GB, and GIB, where each dot is colored by the stationary frequency of cooperation, $\langle x_\mathrm{C} \rangle$, on the corresponding hypergraph, and arrows indicating the increasing directions of the three rules.
    The central region around $1/3$ is linearly rescaled 10-fold for visualization.
    The simplex is partitioned into regions of rule dominance, with percentages showing their respective proportions.
    \textbf{b}, Distributions of the stationary frequency of every imitating rule, $\langle x_u \rangle$ for $u \in \{\mathrm{IB}, \mathrm{GB}, \mathrm{GIB} \}$ for the sampled hypergraphs, shown as box plots (median, interquartile range, and full range).
    The black dashed line depicts the neutral equilibrium, \textit{i.e.}, $\langle x_u \rangle = 1/3$.
    \textbf{c}--\textbf{h}, Impact of varying average orders ($\langle g \rangle$; \textbf{c}--\textbf{e}) and average hyperdegrees ($\langle k \rangle$; \textbf{f}--\textbf{h}) on stationary strategy frequencies for the sampled hypergraphs.
    Ternary plots are shown in a $10\times$ zoomed-in view as well.
    The gray dashed line delineates the neutral equilibrium of either $\langle x_\mathrm{C} \rangle = 1/2$ or $\langle x_u \rangle = 1/3$.
    \textbf{i}--\textbf{k}, Stationary frequencies of cooperation versus distinct imitating rules on the sampled hypergraphs, with dots colored by density.
    Spearman rank correlations of stationary frequencies between cooperation and each rule are shown in the corresponding panel ($\rho_s=-0.769$, $0.754$ and $-0.630$ for IB, GB and GIB, respectively).
    Parameters: $R=1.17$, $\delta = 0.01$, and $v = 0.05$.
    } 
    \label{fig:Fig5}
\end{figure}

\begin{figure}[p]
    \centering
    \includegraphics[width=\linewidth]{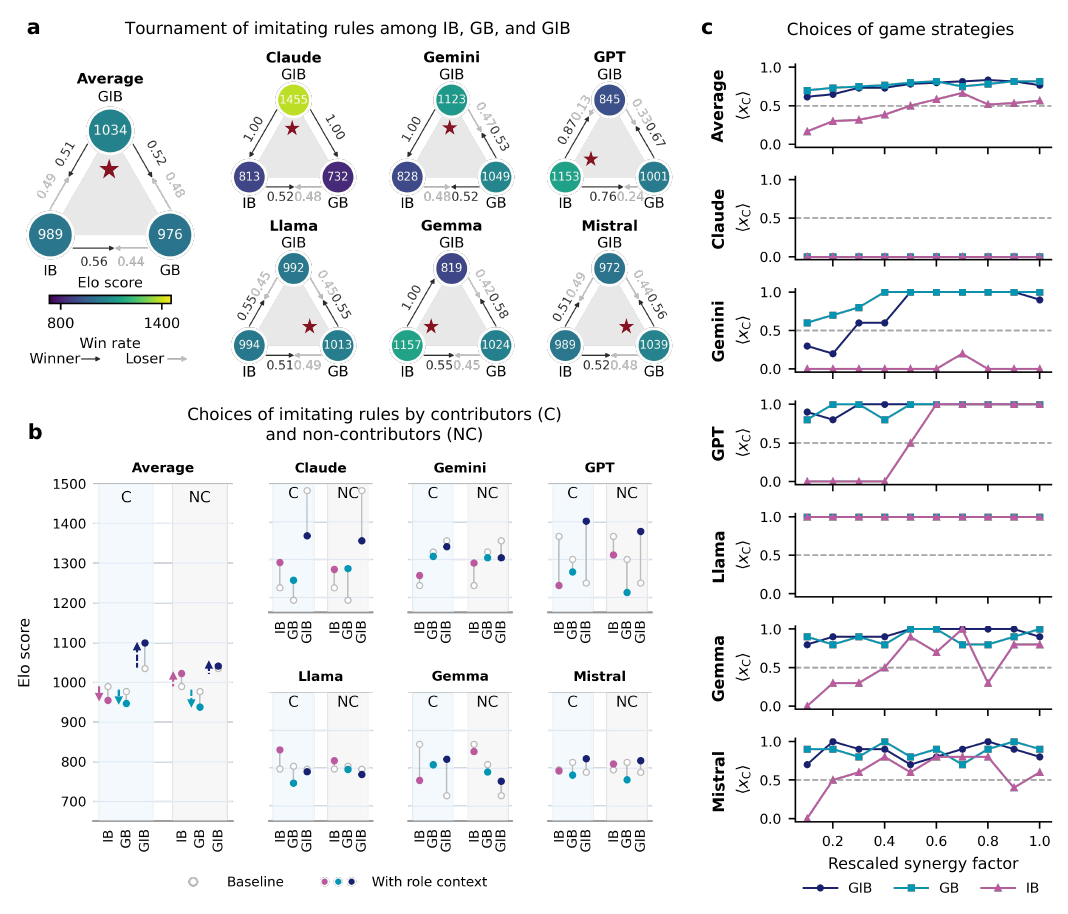}
    \vspace{0.3cm}
    \caption{\textbf{LLM choices of imitating rules and game strategies.}
    Six state-of-the-art Large Language Model (LLM) families are used in the whole figure, including Claude, Gemini, GPT, Llama, Gemma, and Mistral.
    Results are shown for the average outcome of LLM choices and for each individual LLM family separately.
    \textbf{a}, Tournament of imitating rules among IB, GB, and GIB. 
    In each triangle, nodes represent Elo scores of the three imitating rules, with the highest-ranked rule indicated by a star in the tournament;
    directed edges indicate pairwise win rates, pointing from the preferred rule to the less preferred one. 
    \textbf{b}, Elo scores of imitating rules evaluated by contributors (C) and non-contributors (NC).
    Open circles denote baseline Elo scores, colored circles denote Elo scores with role contexts, and dashed arrows indicate within-model shifts from scores in baselines to those with role contexts.
    \textbf{c}, Proportion of choosing contributors, $\langle x_\mathrm{C} \rangle$, in the public goods game setting under different social biases, with regard to rescaled synergy factors.
} 
    \label{fig:Fig6}
\end{figure}

\clearpage

\bibliographystyle{naturemag}
\bibliography{bibliography_v3.bib}

\clearpage

\section*{Methods}
Here we outline the model setup, analytical framework and hypergraph construction below. 
For complete derivations, please refer to Supplementary Information.

\subsection*{Modeling coevolutionary dynamics on hypergraphs}
\noindent \textbf{Population state.} We model a general framework for coevolutionary dynamics of social behaviors and imitating rules in a higher-order population of size $N$.
Here, each individual chooses the social behavior in the public goods game from $\mathcal{S} = \{ \mathrm{C}, \mathrm{D} \}$, and adopts the imitating rule for strategy updating from $\mathcal{U} \subseteq \{ \mathrm{IB}, \mathrm{GB}, \mathrm{GIB} \}$.
The state combination of the population is thus described by a vector $\mathbf{x} = (x_i)_{i=1}^{N}$ with $x_i=(s_i,u_i)$ for $s_i \in \mathcal{S}$ and $u_i \in \mathcal{U}$.
To facilitate the subsequent derivations, we introduce the notation $\mathbbm{1}_{(s',u')}(x_i)$, which equals $1$ for $s_i = s'$ and $u_i = u'$ and $0$ otherwise, for any $s' \in \mathcal{S}$ and $u' \in \mathcal{U}$.

\vspace{0.5cm}
\noindent \textbf{Hypergraph and its properties.} To model the higher-order population structure, we use a hypergraph that consists of a set of $N$ nodes, denoted as $\mathcal{N}$, and a set of $E$ hyperedges, denoted as $\mathcal{E} = \{ I \vert I \subseteq \mathcal{N} \}$.
The hypergraph can be mathematically represented as an incidence indicator $b(i, I)$, which equals $1$ if node $i$ belongs to hyperedge $I$, and $0$ otherwise.
Based on this representation, the hyperdegree of node $i$ is then defined as $k_i = \sum_{I \in \mathcal{E}} b(i, I)$,
and the order of hyperedge $I$ is given by $g_I = \sum_{i \in \mathcal{N}} b(i, I)$.
In addition, for a higher-order random walk on the hypergraph, the walker starting from an original node $i$ moves towards node $j$ ($j=i$ is allowed) via any common hyperedge with the probability $p_{ij}=\sum_{I \in \mathcal{E}} q(i,I)q(I,j)$, where $q(i,I) = b(i,I)/k_i$ and $q(I,i) = b(i,I)/g_I$ are the probabilities that $i$ selects $I$ and $i$ selects $j$ inside $I$, respectively.

\vspace{0.5cm}
\noindent \textbf{Public goods game on hypergraph.} Regarding public goods games on the hypergraph,  games are played over hyperedges.
In a hyperedge of size $g$, each of the $n_\mathrm{C}$ cooperators contributes a fixed cost $1$ to the public pool. 
The total contribution is multiplied by the synergy factor $r$ and afterwards equally shared among all participants, yielding cooperators' and defectors' payoffs as
\begin{equation}
    f_{\mathrm{C}} = \frac{n_{\mathrm{C}}r}{g}-1,~~
    f_{\mathrm{D}} = \frac{n_{\mathrm{C}}r}{g}.
\end{equation}
In the hypergraph context, individual $i$ plays public goods games over hyperedges it belongs to, and thereafter aggregates all the payoffs to obtain its final payoff, that is
\begin{equation}
    f_i(\mathbf{x}) 
   = \sum_{u' \in \mathcal{U}} \left(  \sum_{l \in \mathcal{N}} B_{li} \mathbbm{1}_{(\mathrm{C}, u')}(x_l) r - C_i\mathbbm{1}_{(\mathrm{C}, u')}(x_i) \right),
   \label{payoff}
\end{equation}
where $B_{li} = \sum_{I \in \mathcal{E}} b(i,I) q(I,l)$ is the benefit weight from $l$ to $i$, so that $B_{li}r$ gives the benefit that $i$ receives from a cooperator $l$; and $C_i = k_i$ is the total cost paid by $i$ when cooperating.
Furthermore, the group payoff is defined as the average payoff of group members, $f_I(\mathbf{x}) = \sum_{i \in \mathcal{N}} q(I, i) f_i(\mathbf{x})$.
Payoff is then converted into fitness by $F=e^{\delta f}$, with the selection strength $0 < \delta \ll 1$.
Unless otherwise stated, theoretical analyses in Methods assume weak selection.

\vspace{0.5cm}
\noindent \textbf{Asynchronous strategy update.} 
In each time step, a focal individual is uniformly-at-random determined to update its strategy, either by mutation with probability $v$ or by imitation otherwise.
When imitating, individual $j$ learns the strategy combination from individual $i$ with probability
\begin{equation}
    e_{ij}(\mathbf{x}) = 
    \sum_{u' \in \mathcal{U}} \sum_{s' \in \mathcal{S}} \mathbbm{1}_{(s', u')}(x_j)  e_{ij}^{u'}(\mathbf{x}).
\end{equation}

\begin{equation}
    e_{ij}(\mathbf{x})=  
    \begin{cases}
    e_{ij}^{\mathrm{IB}}(\mathbf{x}) & j \sim \mathrm{IB} \\
    e_{ij}^{\mathrm{GB}}(\mathbf{x}) & j \sim \mathrm{GB} \\
    e_{ij}^{\mathrm{GIB}}(\mathbf{x}) & j \sim \mathrm{GIB} \\
    \end{cases}
\end{equation}

The transition probabilities associated with the imitation rules are specified in Eqs~(\ref{rule-ib})--(\ref{rule-gib}).

\subsection*{General framework for the mutation-selection equilibrium}
We develop a mathematical framework for the mutation-selection equilibrium of all potential strategy combinations on any unweighted, connected hypergraph, which is conditioned on weak selection ($\delta \ll 1$) and applicable to any single or multiple coevolving imitating rules.
The stationary frequency of each strategy combination, $\langle x_{(s^*, u^*)} \rangle$, is determined by three crucial factors:
1) mutation-weighted reproductive value;
2) marginal effect of selection on evolution;
3) strategy assortment on neutrality.
expected strategy distribution on any three nodes.

\vspace{0.5cm}
\noindent \textbf{Mutation-weighted reproductive value.}~
The mutation-weighted reproductive value of the individual $i$ for $i \in \mathcal{N}$ is given by summing the probabilities across the timespan and over the whole population.
\begin{equation}
    \pi_i^{\text{mut}} =\sum_{j \in \mathcal{N}} v \left( (\mathbf{I} - (1-v)\mathbf{P})^{-1}\right)_{ji}.
    \label{mut_wei}
\end{equation}

The mutation-selthe \textbf{mutation-weighted reproductive value} characterizes the contribution of an individual to future generations under neutralityection equilibrium is

We obtain
\begin{equation}
    \langle x_{(s^*,u^*)} \rangle = \frac{1}{2|\mathcal{U}|} + \frac{\delta (1-v)}{Nv} \sum_{u' \in \mathcal{U}} \left( \beta^{u'}_{(s^*,u^*)} r - \gamma^{u'}_{(s^*,u^*)} \right) + \mathcal{O}(\delta^2),
    \label{mean_fre_ibs}
\end{equation}
where $\beta^{u'}_{(s^*,u^*)}$ represents the benefit-related terms for the coevolving imitating rule $u' \in \mathcal{U}$, correspondingly given by
\begin{align}
\beta^{\mathrm{GIB}}_{(s^*,u^*)}
 & = \sum_{\substack{s' \in \mathcal{S} \\ u'' \in \mathcal{U}}}
   \sum_{i,j,l \in \mathcal{N}} \pi_j^{\mathrm{mut}}
\Bigg(
p_{ji} B_{li}
- p_{ji} \sum_{k \in \mathcal{N}} p_{jk} B_{lk}
\Bigg) \\
&\quad \cdot
\bigg\langle
\mathbbm{1}_{(s^*,u^*)}(x_i)
\mathbbm{1}_{(s', \mathrm{GIB})}(x_j)
\mathbbm{1}_{(\mathrm{C}, u'')}(x_l)
\bigg\rangle^\circ,
\allowdisplaybreaks[2] \\[0.8em]
\beta^{\mathrm{GB}}_{(s^*,u^*)}
&= \sum_{\substack{s' \in \mathcal{S} \\ u'' \in \mathcal{U}}}
   \sum_{i,j,k,l \in \mathcal{N}} \pi_j^{\mathrm{mut}}
\Bigg(
\sum_{I \in \mathcal{E}} q(j,I) q(I,i) q(I,k)
- p_{ji} p_{jk} 
\Bigg) B_{lk}
\nonumber \\
&\quad \cdot
\Big\langle
\mathbbm{1}_{(s^*,u^*)}(x_i)
\mathbbm{1}_{(s', \mathrm{GB})}(x_j)
\mathbbm{1}_{(\mathrm{C}, u'')}(x_l)
\Big\rangle^\circ,
\allowdisplaybreaks[2] \\[0.8em]
\beta^{\mathrm{IB}}_{(s^*,u^*)}
&= \sum_{\substack{s' \in \mathcal{S} \\ u'' \in \mathcal{U}}}
   \sum_{i,j,l \in \mathcal{N}} \pi_j^{\mathrm{mut}}
\Bigg(
p_{ji} B_{li}
- \sum_{I \in \mathcal{E}} q(j,I) q(I,i)
  \sum_{k \in \mathcal{N}} q(I,k) B_{lk}
\Bigg)
\nonumber \\
&\quad \cdot
\Big\langle
\mathbbm{1}_{(s^*,u^*)}(x_i)
\mathbbm{1}_{(s', \mathrm{IB})}(x_j)
\mathbbm{1}_{(\mathrm{C}, u'')}(x_l)
\Big\rangle^\circ .
\end{align}
and $\gamma^{u'}_{(s^*,u^*)}$ is the cost-related formula for the coevolving imitating rule $u' \in \mathcal{U}$, respectively specified as
\begin{align}
\gamma^{\mathrm{GIB}}_{(s^*,u^*)}
&= \sum_{\substack{s' \in \mathcal{S} \\ u'' \in \mathcal{U}}} \sum_{i,j \in \mathcal{N}} \pi_j^{\mathrm{mut}} p_{ji}
\Bigg(
 C_i
\Big\langle
\mathbbm{1}_{(s^*,u^*)}(x_i)
\mathbbm{1}_{(s', \mathrm{GIB})}(x_j)
\mathbbm{1}_{(\mathrm{C}, u'')}(x_i)
\Big\rangle^\circ
\nonumber \\*
&\quad
- \sum_{k \in \mathcal{N}} p_{jk} C_k
\Big\langle
\mathbbm{1}_{(s^*,u^*)}(x_i)
\mathbbm{1}_{(s', \mathrm{GIB})}(x_j)
\mathbbm{1}_{(\mathrm{C}, u'')}(x_k)
\Big\rangle^\circ
\Bigg),
\allowdisplaybreaks[2] \\[0.8em]
\gamma^{\mathrm{GB}}_{(s^*,u^*)}
&= \sum_{\substack{s' \in \mathcal{S} \\ u'' \in \mathcal{U}}} \sum_{i,j \in \mathcal{N}} \pi_j^{\mathrm{mut}}
\Bigg(
\sum_{I \in \mathcal{E}} q(j,I) q(I,i)
\sum_{k \in \mathcal{N}} q(I,k) C_k
- p_{ji} \sum_{k \in \mathcal{N}} p_{jk} C_k
\Bigg)
\nonumber \\*
&\quad \cdot
\Big\langle
\mathbbm{1}_{(s^*,u^*)}(x_i)
\mathbbm{1}_{(s', \mathrm{GB})}(x_j)
\mathbbm{1}_{(\mathrm{C}, u'')}(x_k)
\Big\rangle^\circ,
\allowdisplaybreaks[2] \\[0.8em]
\gamma^{\mathrm{IB}}_{(s^*,u^*)}
&= \sum_{\substack{s' \in \mathcal{S} \\ u'' \in \mathcal{U}}} \sum_{i,j \in \mathcal{N}} \pi_j^{\mathrm{mut}}
\Bigg(
p_{ji} C_i
\Big\langle
\mathbbm{1}_{(s^*,u^*)}(x_i)
\mathbbm{1}_{(s', \mathrm{IB})}(x_j)
\mathbbm{1}_{(\mathrm{C}, u'')}(x_i)
\Big\rangle^\circ
\nonumber \\*
&\quad
- \sum_{I \in \mathcal{E}} q(j,I) q(I,i)
\sum_{k \in \mathcal{N}} q(I,k) C_k
\Big\langle
\mathbbm{1}_{(s^*,u^*)}(x_i)
\mathbbm{1}_{(s', \mathrm{IB})}(x_j)
\mathbbm{1}_{(\mathrm{C}, u'')}(x_k)
\Big\rangle^\circ
\Bigg).
\end{align}

\vspace{0.5cm}
\noindent \textbf{Identity by state on neutrality.}~This equation (for $i \neq j$) can be transformed into\begin{equation}
    \phi_{ij} = \frac{v}{2|\mathcal{U}|} + \frac{1-v}{2} \sum_{\iota \in \mathcal{N}} \left(  p_{i \iota} \phi_{\iota j} + p_{j \iota} \phi_{i \iota} \right).
    \label{phi_2}
\end{equation}

This equation (for $i \neq j \neq k \neq i$) can be transformed into
\begin{equation}
        \phi_{ijk} =  \frac{v}{6|\mathcal{U}|}\left( \phi_{jk} + \phi_{ik} + \phi_{ij} \right) + \frac{1-v}{3} \sum_{\iota \in \mathcal{N}}  \left(  p_{i \iota} \phi_{\iota j k}  + p_{j \iota} \phi_{i \iota k}
        +  p_{k \iota} \phi_{i j \iota} \right) .
    \label{phi_3}
\end{equation}

\subsection*{Stationary frequency of cooperation}
We offer the formulas of equilibrium frequencies of cooperation on the hypergraph under each single rule as well as coevolving rules below.
The details of mathematical deduction can be found in Supplementary Information.

\vspace{0.5cm}
\noindent \textbf{IB rule.}~Under IB rule, the stationary frequency of cooperation on the hypergraph is equal to
\begin{equation}
\langle x_{\mathrm C}^{\mathrm{IB}} \rangle
=
\frac{1}{2}
+
\frac{\delta(1-v)}{2Nv}
\left(
\beta_{\mathrm C}^{\mathrm{IB}} r
-
\gamma_{\mathrm C}^{\mathrm{IB}}
\right)
+
\mathcal O(\delta^2),
\end{equation}
with
\begin{align*}
\beta_{\mathrm C}^{\mathrm{IB}}
&=
\sum_{i,j,l\in\mathcal N}
\pi_j^{\mathrm{mut}}\left( p_{ji}
B_{li} 
-
\sum_{k\in\mathcal N} \sum_{I\in\mathcal E} q(j,I)q(I,i)q(I,k)
B_{lk}\right) \phi_{il},
\\[0.8em]
\gamma_{\mathrm C}^{\mathrm{IB}}
&=
\sum_{i,j\in\mathcal N}
\pi_j^{\mathrm{mut}}
\left(
p_{ji}C_i
-
\sum_{k\in\mathcal N}
\sum_{I\in\mathcal E} q(j,I)q(I,i)q(I,k)
C_k\,\phi_{ik} \right).
\end{align*}
Accordingly, cooperation is dominated over defection on the hypergraph under IB rule when
\begin{equation}
    \langle x_{\mathrm C}^{\mathrm{IB}} \rangle > \frac{1}{2}
    ~~\Leftrightarrow~~
    R>R_{\mathrm{IB}}^{*}
    :=
    \frac{\gamma_{\mathrm C}^{\mathrm{IB}}}{\langle g \rangle \cdot \beta_{\mathrm C}^{\mathrm{IB}}}.
\end{equation}
where $R$ is the average-order-rescaled synergy factor for the public goods game to ensure the fairness of distinct orders of hyperedges on the hypergraph.

\vspace{0.5cm}
\noindent \textbf{GB rule.}~Under GB rule, the stationary frequency of cooperation on the hypergraph is equal to
\begin{equation}
\langle x_{\mathrm C}^{\mathrm{GB}} \rangle
=
\frac{1}{2}
+
\frac{\delta(1-v)}{2Nv}
\left(
\beta_{\mathrm C}^{\mathrm{GB}} r
-
\gamma_{\mathrm C}^{\mathrm{GB}}
\right)
+
\mathcal O(\delta^2),
\end{equation}
with
\begin{align*}
\beta_{\mathrm C}^{\mathrm{GB}}
&=
\sum_{i,j,k,l\in\mathcal N}
\pi_j^{\mathrm{mut}}
\left( \sum_{I\in\mathcal E} q(j,I)q(I,i)q(I,k) - p_{ji}p_{jk} \right)
B_{lk}\phi_{il},
\\[0.8em]
\gamma_{\mathrm C}^{\mathrm{GB}}
&=
\sum_{i,j,k\in\mathcal N}
\pi_j^{\mathrm{mut}}
\left( \sum_{I\in\mathcal E} q(j,I)q(I,i)q(I,k) - p_{ji}p_{jk} \right)
C_k\phi_{ik}.
\end{align*}
Accordingly, cooperation is dominated over defection on the hypergraph under GB rule when
\begin{equation}
    \langle x_{\mathrm C}^{\mathrm{GB}} \rangle > \frac{1}{2}
    ~~\Leftrightarrow~~
    R>R_{\mathrm{GB}}^{*}
    :=
    \frac{\gamma_{\mathrm C}^{\mathrm{GB}}}{\langle g \rangle \cdot \beta_{\mathrm C}^{\mathrm{GB}}}.
\end{equation}

\vspace{0.5cm}
\noindent \textbf{GIB rule.}~Under GIB rule, the stationary frequency of cooperation on the hypergraph is equal to
\begin{equation}
\langle x_{\mathrm C}^{\mathrm{GIB}} \rangle
=
\frac{1}{2}
+
\frac{\delta(1-v)}{2Nv}
\left(
\beta_{\mathrm C}^{\mathrm{GIB}} r
-
\gamma_{\mathrm C}^{\mathrm{GIB}}
\right)
+
\mathcal O(\delta^2),
\end{equation}
with
\begin{align*}
\beta_{\mathrm C}^{\mathrm{GIB}}
&=
\sum_{i,j,l\in\mathcal N}
\pi_j^{\mathrm{mut}} p_{ji}
\left( B_{li}
- \sum_{k \in\mathcal N} p_{jk}
B_{lk} \right)\phi_{il};
\\[0.8em]
\gamma_{\mathrm C}^{\mathrm{GIB}}
&=
\sum_{i,j\in\mathcal N}
\pi_j^{\mathrm{mut}} p_{ji} \left( C_i
- \sum_{k \in\mathcal N} p_{jk}C_k\phi_{ik} \right).
\end{align*}
Accordingly, cooperation is dominated over defection on the hypergraph under GIB rule when
\begin{equation}
    \langle x_{\mathrm C}^{\mathrm{GIB}} \rangle > \frac{1}{2}
    ~~\Leftrightarrow~~
    R>R_{\mathrm{GIB}}^{*}
    :=
    \frac{\gamma_{\mathrm C}^{\mathrm{GIB}}}{\langle g \rangle \cdot \beta_{\mathrm C}^{\mathrm{GIB}}}.
\end{equation}

\vspace{0.5cm}
\noindent \textbf{IB, GB, and GIB coevolving rules.} Under all the three coevolving rules, the stationary frequency of cooperation on the hypergraph is equal to
\begin{equation}
\begin{aligned}
\langle x_{\mathrm{C}} \rangle 
= \frac{1}{2}
+ \frac{\delta (1-v)}{Nv}
\left( \beta_{\mathrm{C}} r - \gamma_{\mathrm{C}} \right)
+ \mathcal{O}(\delta^2),
\end{aligned}
\end{equation}
with
\begin{align*}
\beta_{\mathrm{C}}
&=
\sum_{i,j,l\in\mathcal{N}} \pi_j^{\mathrm{mut}}p_{ji}
\bigg(
B_{li}
-
\sum_{k\in\mathcal{N}} p_{jk}B_{lk}
\bigg)
\left(
\frac{1}{5}\phi_{il}+\frac{2}{15}
\right);
\\[0.8em]
\gamma_{\mathrm{C}}
&=
\frac{1}{3} \sum_{i,j\in\mathcal{N}} \pi_j^{\mathrm{mut}} p_{ji} C_i
-
\sum_{i,j,k\in\mathcal{N}} \pi_j^{\mathrm{mut}} p_{ji} p_{jk} C_k
\left(
\frac{1}{5}\phi_{ik}+\frac{2}{15}
\right)
\\
&\quad
-
\sum_{i,j,k \in\mathcal{N}} \pi_j^{\mathrm{mut}}
\sum_{I\in\mathcal{E}} q(j,I) q(I,i) q(I,k) C_k
\left(
\frac{3}{10}\phi_{ijk}
+\frac{3}{20}\phi_{ij}
-\frac{1}{20}\phi_{ik}
-\frac{1}{20}\phi_{jk}
-\frac{1}{60}
\right).
\end{align*}
Accordingly, cooperation is dominated over defection on the hypergraph under the coevolving rules when
\begin{equation}
    \langle x_{\mathrm C} \rangle > \frac{1}{2}
    ~~\Leftrightarrow~~
    R>R^{*}
    :=
    \frac{\gamma_{\mathrm C}}{\langle g \rangle \cdot \beta_{\mathrm C}}.
\end{equation}

\subsection*{Well-mixed hypergraph}
For a well-mixed hypergraph of size $N$, consisting of a single hyperedge that encompasses all nodes, we derive the closed-form expression for the mutation-selection equilibrium of every strategy combination under three coevolving rules.
The stationary frequency of cooperation is given by
\begin{equation}
\langle x_{\mathrm{C}} \rangle =
\frac{1}{2}
-
\frac{\delta(1-v)(N-1)}{6\bigl(Nv-v+1\bigr)}. 
\end{equation}
The corresponding rule-specific components are
\begin{align*}
    &\langle x_{(\mathrm{C}, \mathrm{IB})} \rangle  = \langle x_{(\mathrm{C}, \mathrm{GIB})}\rangle 
     = \frac{1}{6} -  \frac{\delta(1-v) (N-1)\bigl(2Nv - 4v + 5\bigr)}{36\bigl(Nv - 2v + 2\bigr)\bigl(Nv - v + 1\bigr)} ,
    \allowdisplaybreaks[2] \\[0.8em]
    & \langle x_{(\mathrm{C}, \mathrm{GB})} \rangle  = \frac{1}{6} -  \frac{\delta (1-v)\left(N - 1\right) \left(N v - 2 v + 1\right)}
     {18(N v - 2 v + 2)(N v - v + 1)}.
\end{align*}
The mutation-selection equilibrium over the full strategy space is completely reported in Section 4.1.1 of the Supplementary Information.

\subsection*{Synthetic hypergraph investigations}
In Fig.~\ref{fig:Fig5}, we compute the mean strategy frequencies under distinct rules for four families of hypergraph models.
Synthetic hypergraphs are initialized with size $N=100$; to ensure connectivity, we retain only the largest connected component of each random hypergraph.
For every hypergraph family, parameters are sampled uniformly from the ranges specified below.

\vspace{0.5cm}
\noindent \textbf{Random uniform hypergraph (RUH).}~We sample $555$ random uniform hypergraphs, each characterized by a single hyperedge size (order).
This is realized by drawing hyperedges uniformly at random from all possible $g$-subsets of the node set.
The hyperedge order is varied over $g \in [2,10]$, and the number of hyperedges over $E \in [100,160]$.

\vspace{0.5cm}
\noindent \textbf{Random non-uniform hypergraph (RNH).}~We generate $1,452$ random non-uniform hypergraphs in which hyperedges of multiple orders can coexist.
Hyperedge orders are sampled either from a shifted Poisson distribution $g = 2 + \mathrm{Poisson}(\lambda)$ with $\lambda \in [0.1, 18]$, or from a truncated power-law distribution $P(g) \propto g^{-\alpha}$ with $\alpha \in [1.2, 3.0]$. 
The total number of hyperedges ranges from $E=100$ to $160$. 
The resulting dataset contains $537$ Poisson RNHs and $915$ power-law RNHs, constructed by uniformly selecting node subsets of the prescribed sizes as hyperedges.

\vspace{0.5cm}
\noindent \textbf{Scale-free hypergraph (SFH).}~We construct $13,451$ scale-free hypergraphs in which hyperedges have their own sizes and node participation in different sizes of hyperedges follows a hidden-parameter model.
Hyperedge orders are sampled from a shifted Poisson distribution $g = 2 + \mathrm{Poisson}(\lambda)$ with $\lambda \in [0.5, 18]$. 
The total number of hyperedges ranges from $E=50$ to $160$. 
The hidden-parameter scale controlling heterogeneity in node participation ranges from $0.8$ to $2.4$. 
As a result, hypergraphs are constructed by preferentially selecting node subsets of the prescribed sizes as hyperedges.

\vspace{0.5cm}
\noindent \textbf{Community hypergraph (ComH).}~We build up $1,011$ community hypergraphs using hypergraph mixed-membership stochastic block model proposed by Ruggeri \textit{et al.}~\cite{nicolo2024}.
Specifically, we consider the number of communities ranging from $2$ to $10$ and the average hyperdegree from $2$ to $10$. 
The parameter $\alpha \in [0.2, 1.0]$ controls how nodes distribute their affiliations across different communities, and $\varepsilon \in [0.1, 0.9]$ governs the strength of community structure in the hypergraph.

\vspace{0.5cm}
For these random hypergraphs, theoretical mutation-selection equilibria are computed using Eq.~\ref{mean_fre_ibs}, without numerical simulations.
Additionally, we explore the impact of hypergraph properties on rescaled critical synergy factors of cooperation in Extended data Fig.~\ref{fig:FigExt8}.
Complementary results for the same hypergraph families with size $30$ are shown in Fig.~\ref{fig:Fig2}d–f, Supplementary Fig.~3, and Supplementary Fig.~5.

\subsection*{Empirical higher-order populations}
To better understand evolutionary mechanisms in real-world systems, we analyze structural properties and evolutionary outcomes on 24 empirical higher-order populations across seven categories.
From each population, we extract a connected subpopulation and construct the corresponding higher-order network.
For these networks, evolutionary outcomes of theoretical predictions are investigated in Fig.~\ref{fig:Fig2}g--i and Supplementary Fig.~2;
illustrations and summary statistics can be found in Fig.~\ref{fig:Fig2}a as well as Extended data Fig.~\ref{fig:FigExt3};
and their descriptions are detailed below.

\begin{itemize}
    \item {Coauthorship [DBLP~\cite{Benson2018}, Geology, History~\cite{Sinha-2015-MAG}]:}~Coauthorship networks encode scientific collaborations, where nodes are authors and hyperedges are joint publications. 
    We consider three coauthorship datasets: 
    DBLP, constructed from publications indexed by DBLP; 
    {Geology}, consisting of publications tagged as “Geology” in the Microsoft Academic Graph; 
    and {History}, consisting of publications tagged as “History” in the Microsoft Academic Graph.
    \item {Political collaboration [Congress bills, Senate bills, House bills, Senate committees, House committees~\cite{Fowler-2006-connecting, Fowler-2006-cosponsorship, chodrow2021hypergraph}]:}~Political networks characterize higher-order interactions among U.S. legislators, where nodes represent Congress members and hyperedges capture collective political activities, including bill co-sponsorships and committee memberships. 
    Specifically, {Congress bills} dataset records temporal simplices formed by sponsors and co-sponsors of legislative bills introduced in both chambers of Congress. 
    {Senate bills} and {House bills} datasets represent co-sponsorship in the Senate and House of Representatives, respectively, while {Senate committees} and {House committees} datasets describe committee membership in the two chambers.
    \item {Human contact [primary school~\cite{Stehl-2011-contact}, high school~\cite{Mastrandrea-2015-contact}, hospital~\cite{Genois-2018-colocation}, village~\cite{Ozella-2021-contact}]:}~Contact networks capture physical proximity among individuals, where nodes are persons and hyperedges represent groups of individuals who were simultaneously close to one another, recorded by wearable or Bluetooth sensors. 
    \item {Online social media [YouTube, Flickr, LiveJournal~\cite{Mislove-2007-social}]:}~Online social networks describe friendship communities among users, where nodes are users and hyperedges represent groups of mutually connected users who also share at least one common community. 
    \item {Email communication [European~\cite{Leskovec-2007-evolution}, Enron~\cite{Benson2018}, W3C~\cite{Craswell-2005-TREC}]:}~Email networks capture group communication through emails, where nodes are email addresses and hyperedges consist of the sender together with all recipients of the same email. 
    We consider three email sources: {European} dataset records email exchanges in an European research institution; {Enron} dataset records emails among a core set of Enron employees; and {W3C} dataset is constructed from W3C mailing lists~.
    \item {Co-review [music~\cite{ni2019justifying}, restaurant, bar~\cite{amburg2020fair}]:}~Co-review networks regard reviewers as nodes and groups of category-specific reviewers as hyperedges. 
    We consider three settings: {music}, based on Amazon blues music reviews; {restaurant}, based on Yelp restaurant reviews in Madison, WI; and {bar}, based on Yelp bar reviews in Las Vegas, NV.
    \item {Q\&A threads [{Stack Overflow}, {Mathematics Stack Exchange}, {Ask Ubuntu}~\cite{Benson2018}]:}~Q\&A thread networks capture the participation in online discussion threads, where nodes are users and hyperedges correspond to groups of users participating in the same thread within a 24-hour time window. 
\end{itemize}

\subsection*{LLM experiments}
We conduct all Large Language Model (LLM) experiments through the Dartmouth Chat API.
Six model families are evaluated, with the following labels used throughout the analyses and figures:
\begin{center}
\small
\begin{tabular}{@{}lll@{}}
\toprule
Label & Model family & Model identifier \\
\midrule
GPT & OpenAI GPT & \texttt{openai.gpt-oss-120b} \\
Gemma & Google Gemma & \texttt{google.gemma-3-27b-it} \\
Claude & Anthropic Claude & \texttt{anthropic.claude-haiku-4-5-20251001} \\
Llama & Meta Llama & \texttt{meta.llama-3-2-3b-instruct} \\
Gemini & Google Gemini & \texttt{vertex\_ai.gemini-3-flash-preview} \\
Mistral & Mistral & \texttt{mistral.mistral-medium-2508} \\
\bottomrule
\end{tabular}
\end{center}
For all models, the sampling temperature is fixed at $0.0$.
For experiments of choosing the imitating rules (Fig.~\ref{fig:Fig6}a--b), each model is asked to decide between two randomly ordered imitating rules drawn from IB, GB and GIB.
Each model completes $300$ pairwise trials in each prompt context.
Specific prompts in all the experiments are provided in Supplementary Information.
We summarize these pairwise tournaments using directed win rates and Elo scores.
For each model and prompt context, the three imitating rules are initialized with the same Elo score of $1,000$.
When rules $i$ and $j$ were compared, the expected score of rule $i$ was
\begin{equation}
    p_i =
    \frac{1}{1+10^{(\mathcal{E}_j-\mathcal{E}_i)/400}},
\end{equation}
with $p_j=1-p_i$.
If the LLM selects rule $i$, we assign the observed scores $s_i=1$ and $s_j=0$; if it selects rule $j$, we assign $s_i=0$ and $s_j=1$.
After each pairwise comparison, scores are updated as
\begin{equation}
    \mathcal{E}_i \leftarrow \mathcal{E}_i + K(s_i-p_i),
    \qquad
    \mathcal{E}_j \leftarrow \mathcal{E}_j + K(s_j-p_j),
\end{equation}
with $K=32$.
The final Elo score of each rule was the rating after all pairwise trials in that context.
For each unordered pair of rules, the directed win rate is computed as the fraction of successful comparisons in which one rule is selected over the other.

\subsection*{Spearman rank correlations}
We use the Spearman rank correlation to quantify whether two variables tend to change in the same direction.
This statistic captures monotonic association rather than linear dependence.
For $n$ paired observations, let $X_i$ and $Y_i$ denote the two variables measured for observation $i$.
The Spearman rank correlation is the Pearson correlation between their rank variables,
\begin{equation}
    \rho_s =
    \frac{
    \sum_{i=1}^{n} \left(R_X(i)-\overline{R_X}\right)
    \left(R_Y(i)-\overline{R_Y}\right)}
    {\sqrt{\sum_{i=1}^{n} \left(R_X(i)-\overline{R_X}\right)^2}
    \sqrt{\sum_{i=1}^{n} \left(R_Y(i)-\overline{R_Y}\right)^2}},
\end{equation}
where $R_X(i)$ and $R_Y(i)$ are the ranks of $X_i$ and $Y_i$, respectively, and $\overline{R_X}$ and $\overline{R_Y}$ denote the mean ranks of the two variables.
The coefficient $\rho_s$ ranges from $-1$ to $1$, with positive and negative values indicating positive and negative monotonic correlations, respectively.

\clearpage

\begin{extendedfigure}[p]
    \centering
    \includegraphics[width=\linewidth]{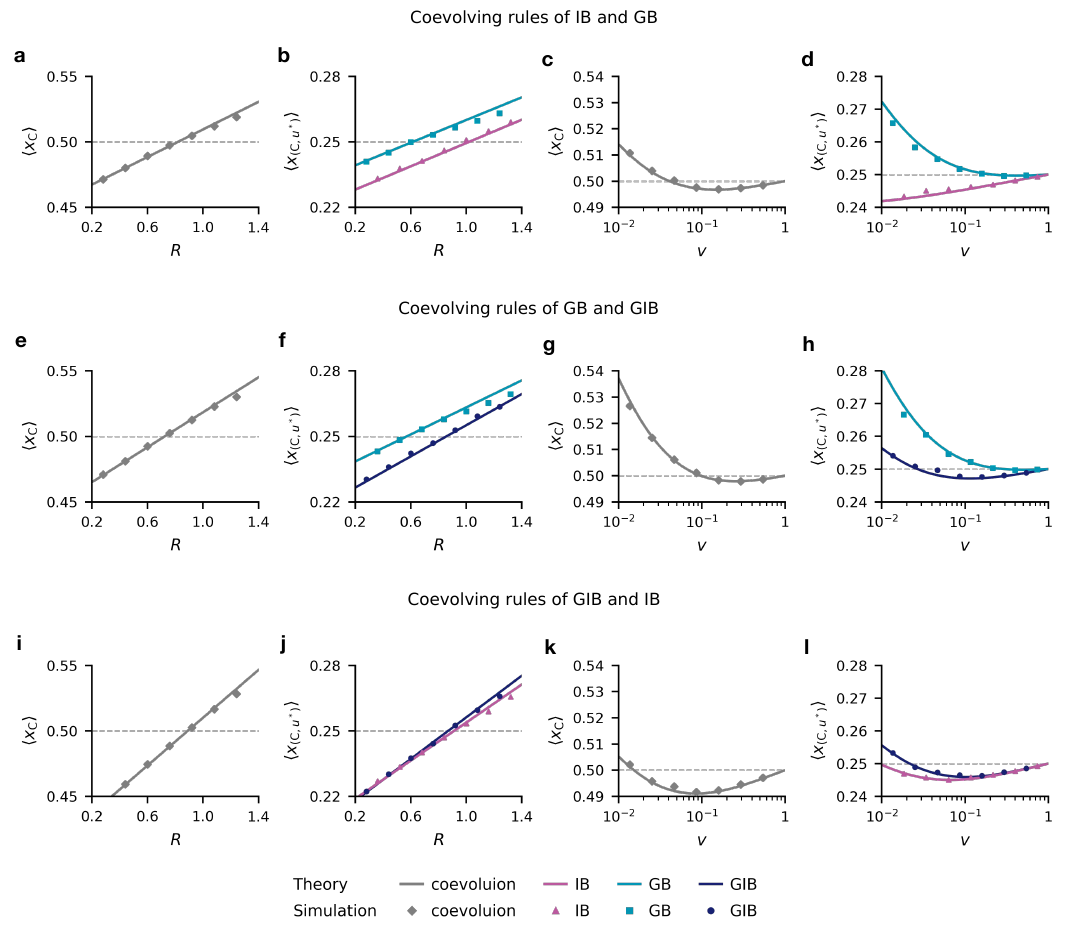}
    \vspace{0.3cm}
    \caption{\textbf{Mean strategy frequencies under every pair of coevolving rules on the coauthorship network.}
    The hypergraph structure is the same as that in Fig.~\ref{fig:Fig2}.
    Panels in the left two columns show mean frequencies of cooperation and its coexistence with each rule as functions of the rescaled synergy factor $R$; 
    and panels in the right two columns show those as functions of the mutation rate $v$.
    Curves represent theoretical predictions, and scatters depict simulation results over the last $10^8$ runs from $10^9$ time steps.
    The gray dashed line indicates the neutrality in each panel.
    Parameters: 
    $\delta=0.025$, 
    $v=0.05$ (\textbf{a}--\textbf{b}, \textbf{e}--\textbf{f}, and \textbf{i}--\textbf{j}), and
    $R=1$ (\textbf{c}--\textbf{d}, \textbf{g}--\textbf{h}, and \textbf{k}--\textbf{l}).
} 
    \label{fig:FigExt1}
\end{extendedfigure}

\clearpage

\begin{extendedfigure}[p]
    \centering
    \includegraphics[width=\linewidth]{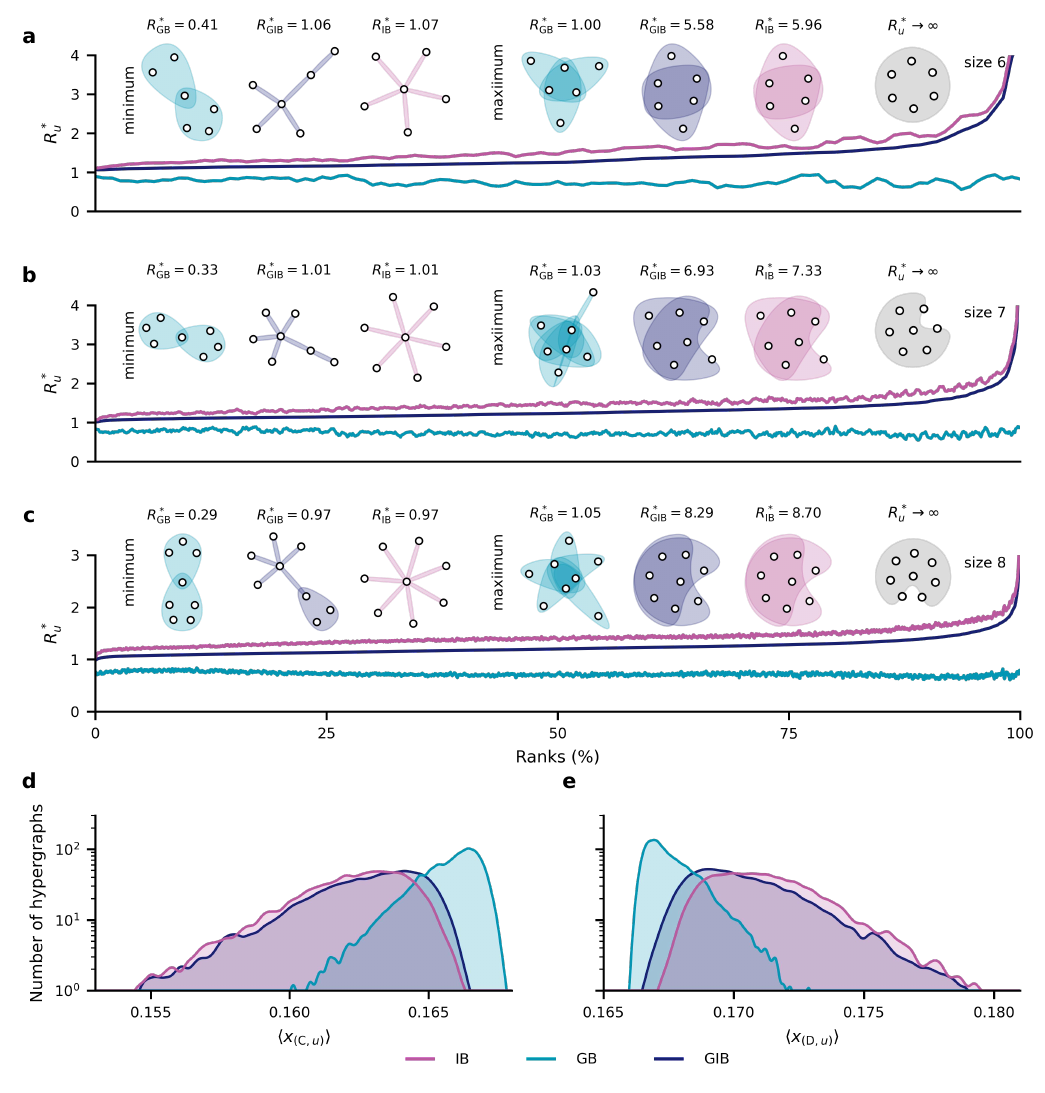}
    \vspace{0.3cm}
    \caption{\textbf{Comparisons of theoretical consequences under distinct rules on 12,082 small-scale hypergraphs.}
    These hypergraphs are generated by treating cliques in underlying pairwise graphs as hyperedges.
    \textbf{a}--\textbf{c}, Ranks of critical thresholds of cooperation under imitating rules (IB, GB, and GIB) across 112 hypergraphs of size 6 (\textbf{a}), 853 hypergraphs of size 7 (\textbf{b}), and 11,117 hypergraphs of size 8 (\textbf{c}), respectively.
    Hypergraphs yielding minimum, maximum (finite), and divergent thresholds of cooperation are visualized separately for each imitating rule.
    \textbf{d}--\textbf{e}, Statistic distributions of mean frequencies of strategy combinations, \textit{i.e.}, $(\mathrm{C}, u)$ for the combination of cooperation and an imitating rule $u \in \{ \mathrm{IB}, \mathrm{GB}, \mathrm{GIB} \}$, and $(\mathrm{D}, u)$ for the combination of defection and an imitating rule $u \in \{ \mathrm{IB}, \mathrm{GB}, \mathrm{GIB} \}$.
    For visual clarity, all ranked curves and statistical distributions are smoothed by binning adjacent hypergraphs.
    Parameters: $v=0.05$ (\textbf{a}--\textbf{e}); $\delta=0.01$ and $R=1.17$ (\textbf{d}--\textbf{e}).
} 
    \label{fig:FigExt2}
\end{extendedfigure}

\clearpage

\begin{extendedfigure}[p]
    \centering
    \includegraphics[width=0.88\linewidth]{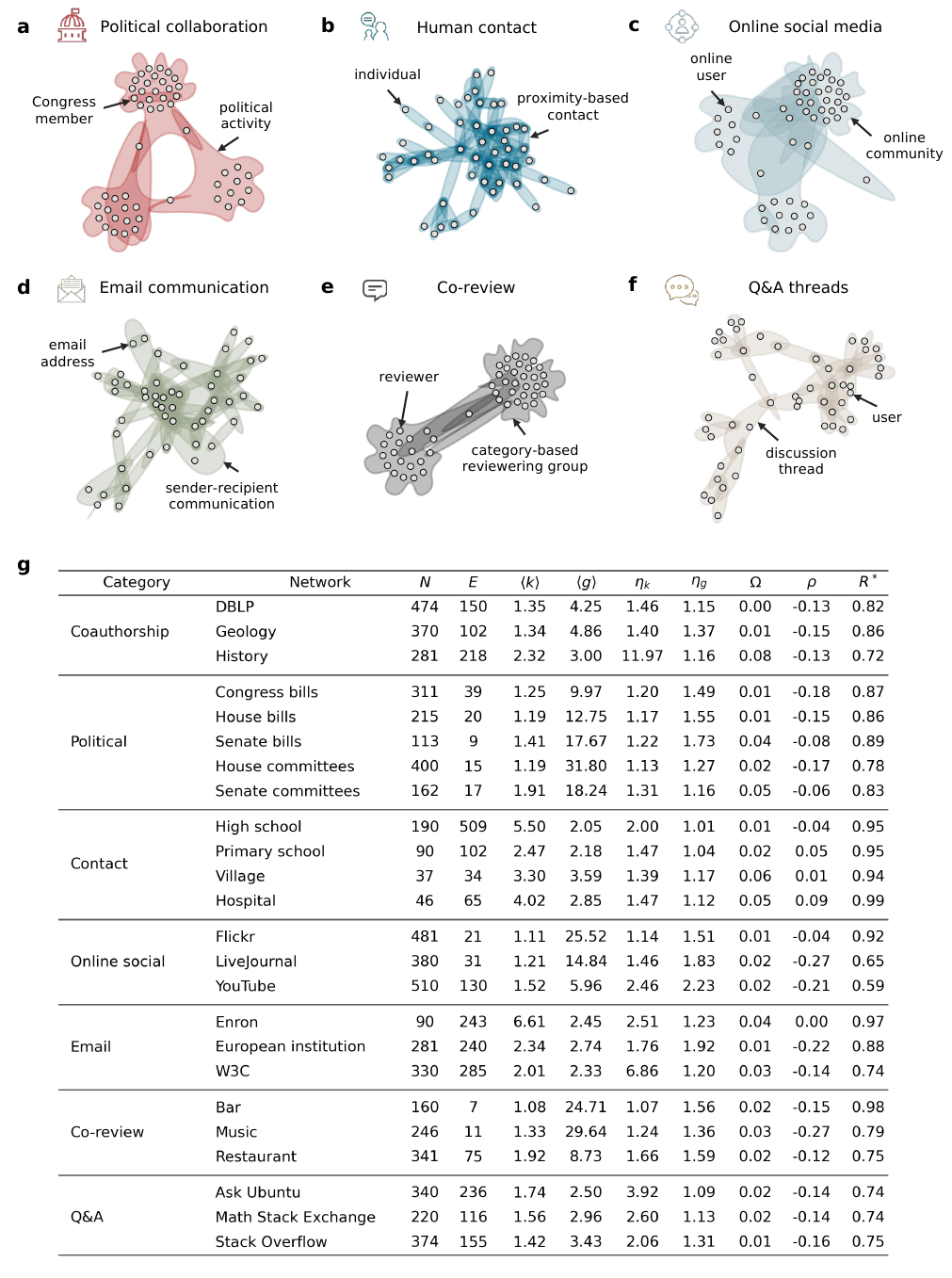}
    \vspace{0.3cm}
    \caption{\textbf{Analyses of empirical higher-order populations.}
    \textbf{a}--\textbf{f}, Illustrations of representative empirical networks from six categories, including political collaboration (\textbf{a}), human contact (\textbf{b}), online social media (\textbf{c}), email communication (\textbf{d}), co-review (\textbf{e}), and Q\&A threads (\textbf{f}).
    A total of 24 empirical hypergraphs from seven categories are analyzed.
    \textbf{g}, Summary statistics of these empirical populations:
    $N$ is the population size; 
    $E$ is the number of hyperedges;
    $\langle k \rangle$ is the average hyperdegree;
    $\langle g \rangle$ is the average order;
    $\eta_k$ (resp. $\eta_g$) is the heterogeneity of the node's hyperdegree (resp. hyperedge's order) distribution;
    $\Omega$ is the strength of overlap among hyperedges;
    $\rho$ is the assortativity between hyperdegrees and orders;
    and $R^*$ is the critical rescaled synergy factor for cooperation under three coevolving rules.
    Definitions of hypergraph properties as well as descriptions of all empirical datasets are provided in Section 1 and 5 of Supplementary Information.
    Parameter: $v=0.05$.
} 
    \label{fig:FigExt3}
\end{extendedfigure}

\clearpage

\begin{extendedfigure}[p]
    \centering
    \includegraphics[width=\linewidth]{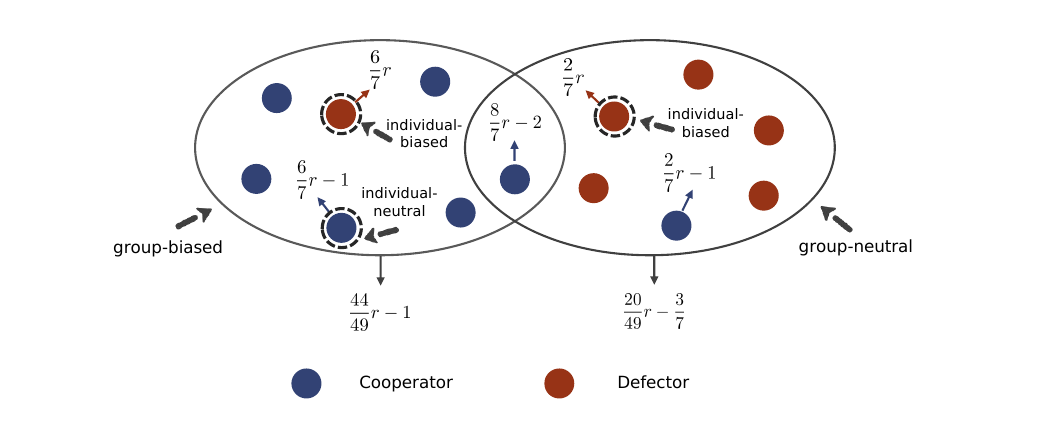}
    \caption{\textbf{Intuitions of selecting a group and selecting an individual with payoff bias or not.}
    The example shows a hypergraph with two hyperedges of size $7$.
    Numbers next to nodes and hyperedges indicate individual payoffs and group average payoffs, respectively, from public goods games on the hypergraph.
    For $r>7/6$, the cooperator-enriched group on the left has a higher group average payoff than the group on the right.
    The central individual therefore tends to select the left group when it is biased to group payoffs, \textit{i.e.}, group-biased; 
    it otherwise assigns equal probability to both groups, \textit{i.e.}, group-neutral.
    In a chosen group, defectors typically receive higher individual payoffs than cooperators because they do not pay the cost.
    Hence, the central individual tends to select a defector for reference when it is biased to individual payoffs, \textit{i.e.}, individual-biased; otherwise, it treats every individual equally, \textit{i.e.}, individual-neutral. 
    This contrast illustrates why a group-biased decision-maker favors cooperators while an individual-biased one favors defectors.
    }
    \label{fig:FigExtss2}
\end{extendedfigure}
    
\clearpage

\begin{extendedfigure}[p]
    \centering
    \includegraphics[width=\linewidth]{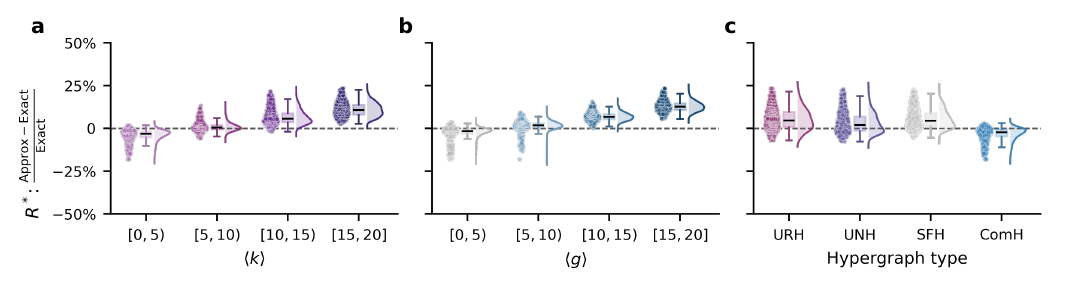}
    \vspace{0.3cm}
    \caption{\textbf{Accuracy validation of mean-field approximations of critical rescaled synergy factors on synthetic hypergraphs.}
    Relative difference between the approximated threshold $R^*$ obtained from Eq.~\ref{gib} and the exact threshold $R^*$ from Fig.~\ref{fig:Fig3}a on synthetic hypergraphs of size 100.
    Results are grouped by average hyperdegree $\langle k \rangle$ (\textbf{a}), average order $\langle g \rangle$ (\textbf{b}), and hypergraph type (\textbf{c}), including random uniform hypergraphs (RUHs), random non-uniform hypergraphs (RNHs), scale-free hypergraphs (SFHs), and community hypergraphs (ComHs).
    Parameter: $v=0.05$.
} 
    \label{fig:FigExt4}
\end{extendedfigure}

\clearpage

\begin{extendedfigure}[p]
    \centering
    \includegraphics[width=\linewidth]{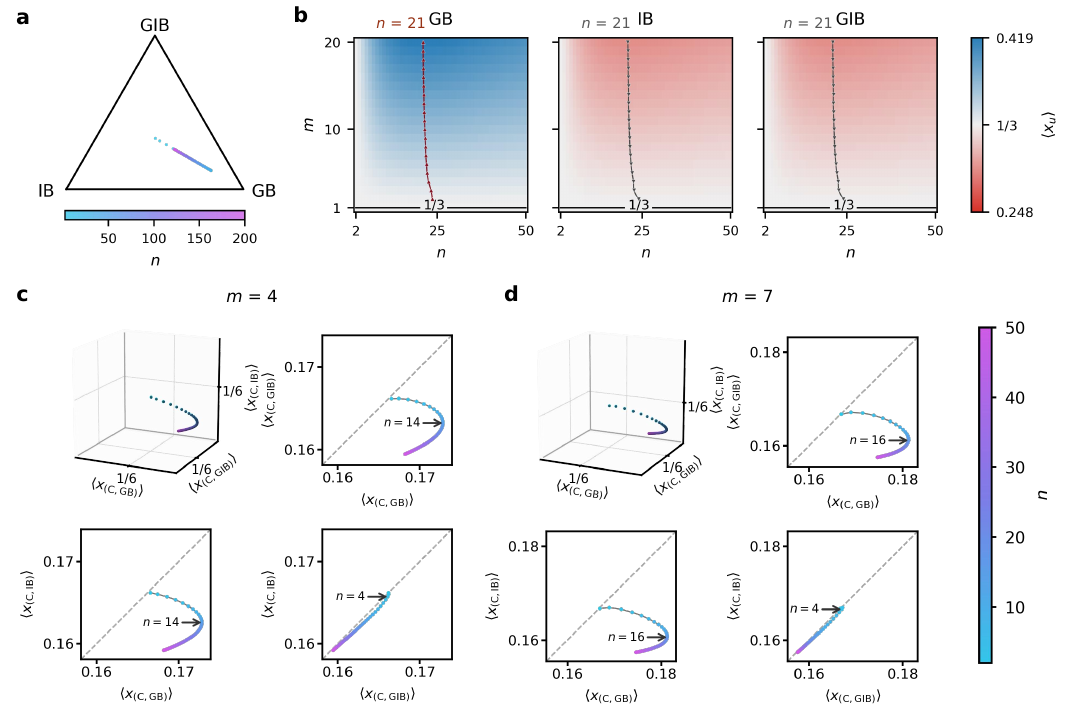}
    \vspace{0.3cm}
    \caption{\textbf{Further results on anchoring-node hypergraphs.}
    \textbf{a}, Evolutionary trajectories of the population composition of imitating rules, shown on a ternary simplex as the group size $n$ varies.
    \textbf{b}, Heat maps of analytical mean frequencies of the three imitating rules, $\langle x_u \rangle$ for $u \in \{ \mathrm{GB},\mathrm{IB},\mathrm{GIB} \}$, over group sizes, $n$, and numbers of peripheral groups, $m$, on anchoring-node hypergraphs.
    The red (resp. gray) line indicates the optimal (resp. pessimal) $n$ for each $m$;
    and the black curves delineate the neutral condition of $\langle x_u \rangle = 1/3$ for every rule $u$.
    \textbf{c}--\textbf{d}, Mean frequencies of combinations of cooperation and distinct rules for anchoring-node hypergraphs with varying $n$ and fixed $m=4$ (\textbf{c}) or $m=7$ (\textbf{d}).
    Arrows annotate trajectory turning points with respect to $n$, at which relative advantages of cooperation between pairs of rules are maximized.
    Parameters: $R=1$, $\delta = 0.025$, and $v = 0.05$.} 
    \label{fig:FigExt5}
\end{extendedfigure}

\clearpage

\begin{extendedfigure}[p]
    \centering
    \includegraphics[width=\linewidth]{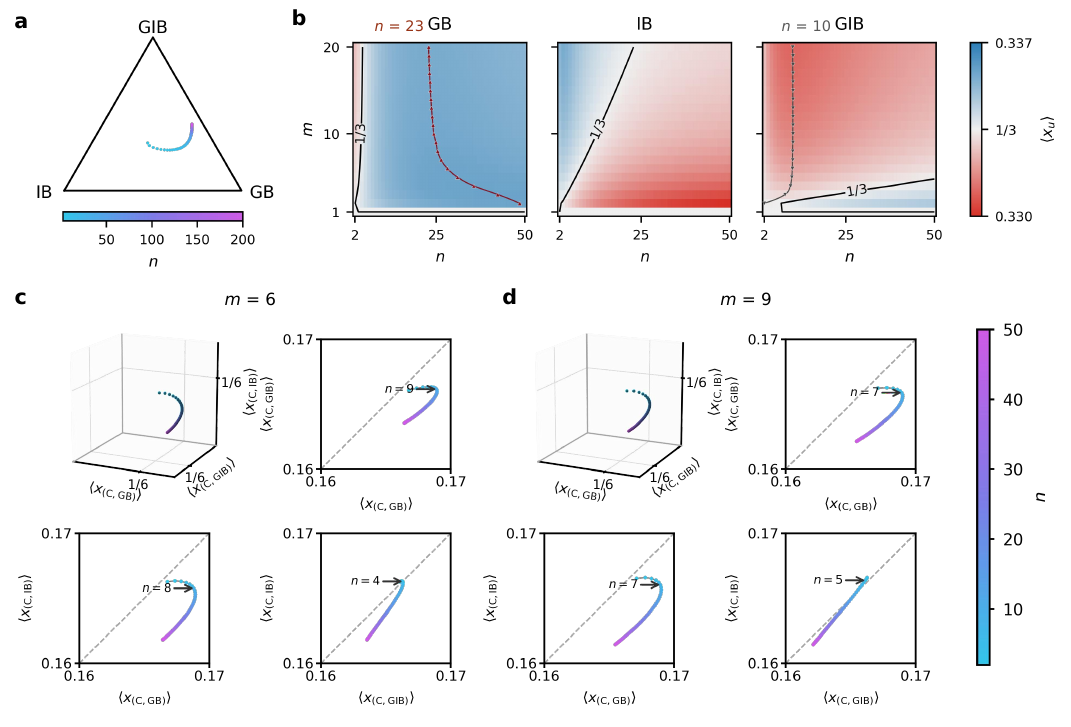}
    \vspace{0.3cm}
    \caption{\textbf{Further results on anchoring-group hypergraphs.}
    All panels and model settings are identical to those in Fig.~\ref{fig:FigExt5}, except that anchoring-group hypergraphs are considered instead of anchoring-node hypergraphs.
    }
    \label{fig:FigExt6}
\end{extendedfigure}

\clearpage

\begin{extendedfigure}[p]
    \centering
    \includegraphics[width=\linewidth]{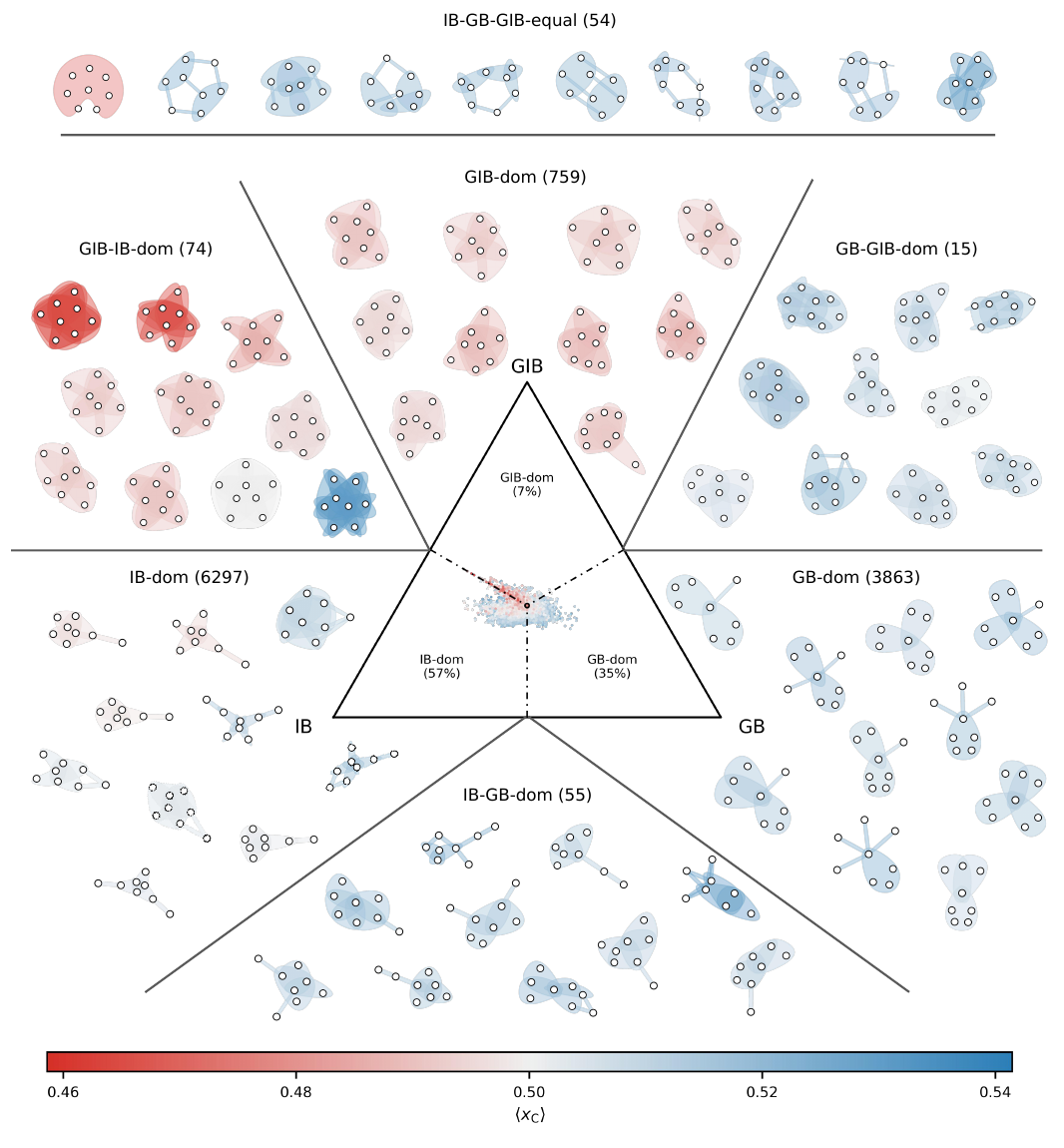}
    \vspace{0.3cm}
    \caption{\textbf{Mean strategy frequencies on 11,117 hypergraphs of size 8.}
    The ternary plot shows the composition of IB, GB, and GIB, with each dot representing a hypergraph and colored by its stationary frequency of cooperation, $\langle x_\mathrm{C} \rangle$.
    Hypergraphs are grouped according to the frequencies of their leading rules, including single-rule dominance (IB-dom, GB-dom, and GIB-dom), tied dominance between two rules (IB-GB-dom, GB-GIB-dom, and GIB-IB-dom), and equal frequencies of all three rules (IB-GB-GIB-equal), with numbers in parentheses denoting the corresponding counts.
    For each regime, ten hypergraphs with the largest dominant-rule frequencies are illustrated.
    Parameters: $R=1.55$, $\delta=0.025$, and $v=0.05$.
} 
    \label{fig:FigExt7}
\end{extendedfigure}

\clearpage

\begin{extendedfigure}[p]
    \centering
    \includegraphics[width=\linewidth]{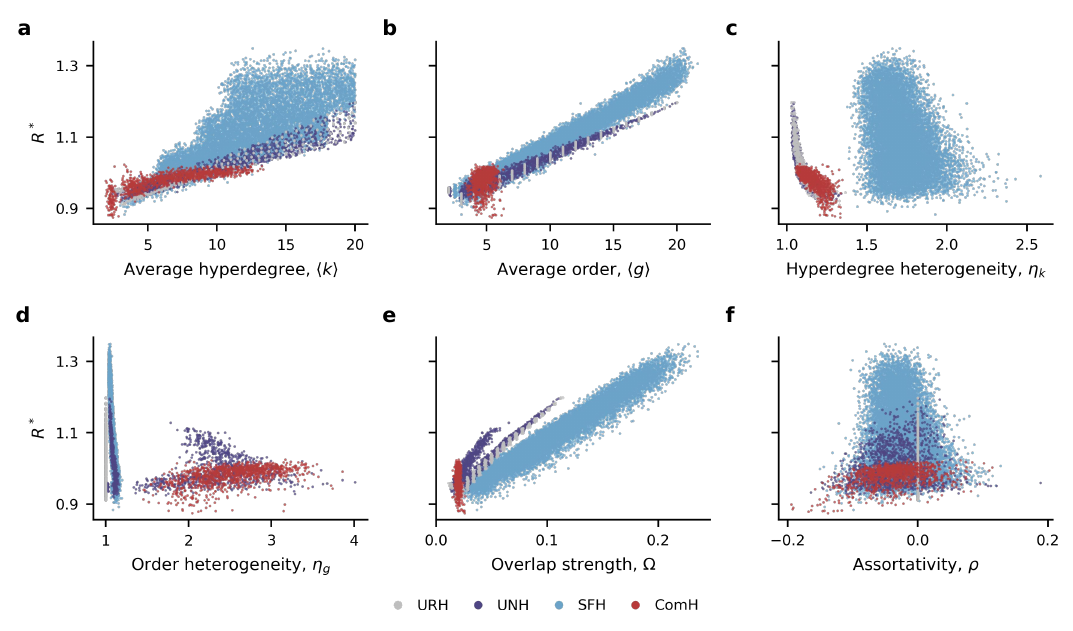}
    \vspace{0.3cm}
    \caption{\textbf{Impact of hypergraph properties on critical thresholds of cooperation under three coevolving rules on synthetic hypergraphs.}
    Scatters of all panels correspond to the same ensemble of synthetic hypergraphs as Fig.~\ref{fig:Fig5}, and hypergraph properties are defined in Supplementary Information.
    Parameters: $v = 0.05$.
} 
    \label{fig:FigExt8}
\end{extendedfigure}

\clearpage

\begin{extendedfigure}[p]
    \centering
    \includegraphics[width=\linewidth]{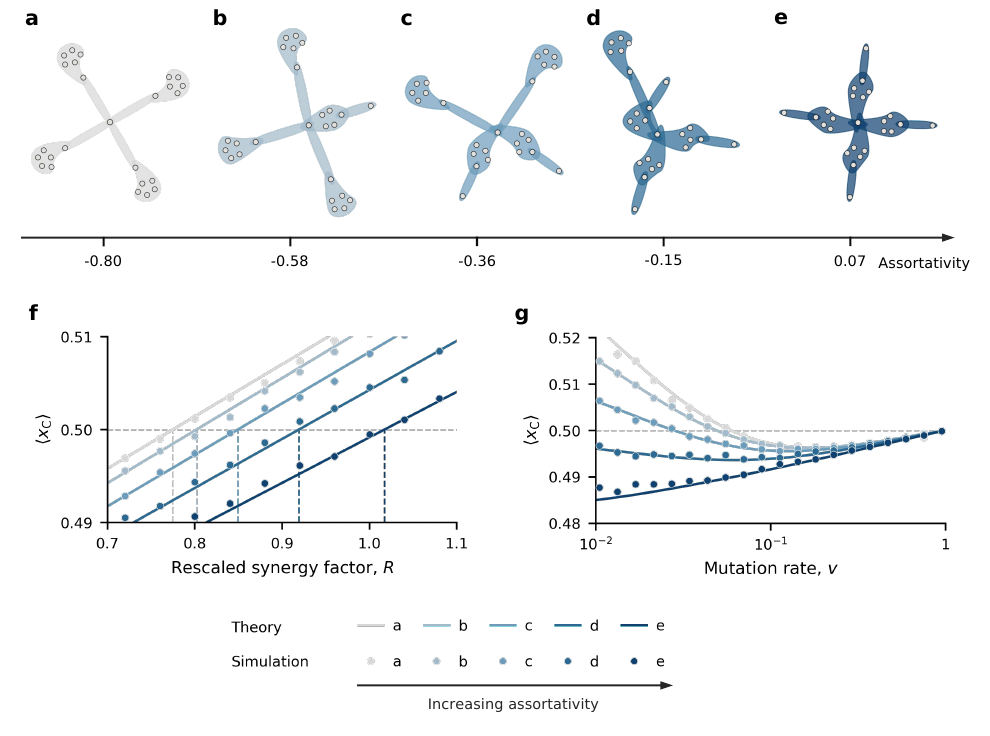}
    \vspace{0.3cm}
    \caption{\textbf{Increasing assortativities between hyperdegrees and orders can foster cooperation under coevolving imitating rules.}
    \textbf{a}--\textbf{e}, Visualizations of hypergraphs with increasing assortativities.
    \textbf{f}--\textbf{g}, Mean frequencies of cooperation under three coevolving rules on hypergraphs \textbf{a}--\textbf{e} as functions of the rescaled synergy factor ($R$; \textbf{f}) as well as the mutation rate ($v$; \textbf{g}).
    Solid lines depict theoretical results, and scatters are from simulation outcomes over the last $10^8$ runs from $10^9$ time steps.
    Parameters: $\delta=0.025$, $v=0.05$ (\textbf{f}), and $R=1.3$ (\textbf{g}).
} 
    \label{fig:FigExt9}
\end{extendedfigure}

\clearpage

\begin{extendedfigure}[p]
    \centering
    \includegraphics[width=\linewidth]{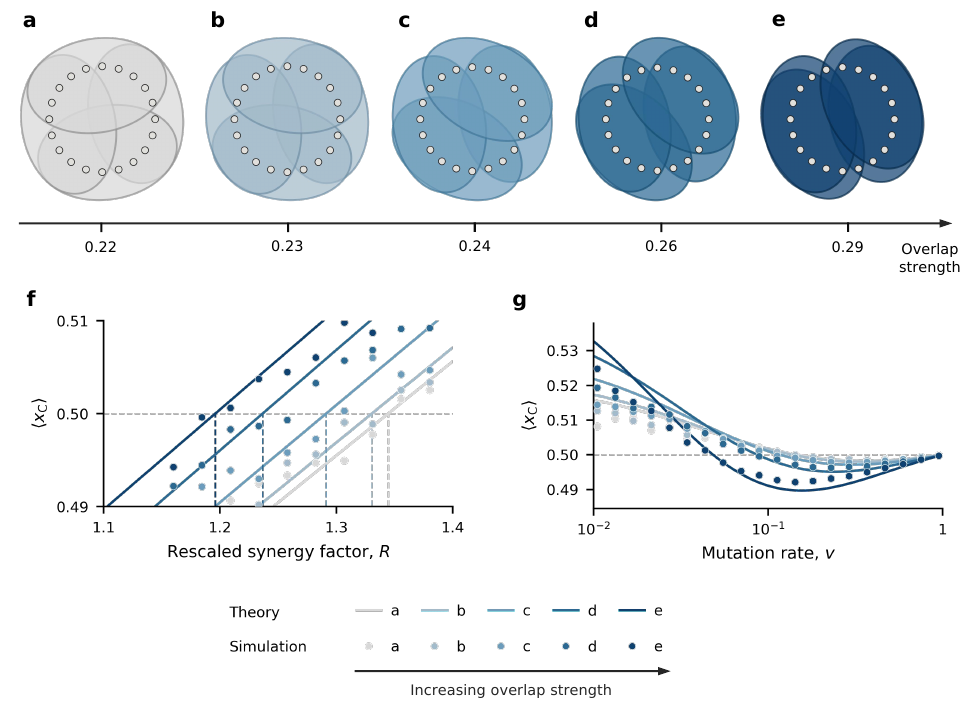}
    \vspace{0.3cm}
    \caption{\textbf{Increasing overlaps among hyperedges can promote cooperation under coevolving imitating rules.}
    Model and parameter settings are the same as those in Fig.~\ref{fig:FigExt9}, except that increasing overlap strengths are considered instead of increasing assortativities between hyperdegrees and orders.
} 
    \label{fig:FigExt10}
\end{extendedfigure}

\end{document}